\documentclass[12pt]{article}

\usepackage{cite}
\usepackage{amsmath}
\usepackage{amssymb}
\usepackage{bm}
\usepackage{color}
\usepackage{graphicx}
\numberwithin{equation}{section}

\DeclareMathOperator{\Det}{Det}

\newcounter{aff}

\begin{document}
\begin{titlepage}
\begin{flushright}
{\footnotesize NITEP 83, OCU-PHYS 523}
\end{flushright}
\begin{center}
{\LARGE\bf
Brane Transitions from Exceptional Groups}\\
\bigskip\bigskip
{\large
Tomohiro Furukawa\footnote{\tt furukawa@sci.osaka-cu.ac.jp},
Sanefumi Moriyama\footnote{\tt moriyama@sci.osaka-cu.ac.jp},
Tomoki Nakanishi\footnote{\tt nakanishi@ka.osaka-cu.ac.jp }
}\\
\bigskip
${}^{*\dagger\ddagger}$\,{\it Department of Physics, Graduate School of Science,}\\
{\it Osaka City University, Sumiyoshi-ku, Osaka 558-8585, Japan}\\[3pt]
${}^{\dagger}$\,{\it Nambu Yoichiro Institute of Theoretical and Experimental Physics (NITEP),}\\
{\it Osaka City University, Sumiyoshi-ku, Osaka 558-8585, Japan}\\[3pt]
${}^\dagger$\,{\it Osaka City University Advanced Mathematical Institute (OCAMI),}\\
{\it Osaka City University, Sumiyoshi-ku, Osaka 558-8585, Japan}
\end{center}

\begin{abstract}
It is a well-known result by Hanany and Witten that, when two five-branes move across each other, D3-branes stretching between them are generated.
Later the same brane configurations played a crucial role in understanding the worldvolume theory of multiple M2-branes.
Recently the partition function of multiple M2-branes was transformed to the Fredholm determinant for quantum algebraic curves, where the characteristic 3/2 power law of degrees of freedom is reproduced and the determinant enjoys a large symmetry given by exceptional Weyl groups.
The large exceptional Weyl group reproduces the Hanany-Witten brane transitions and, besides, contains brane transitions unknown previously.
Aiming at understanding the new brane transitions better, we generalize our previous study on the $D_5$ quantum curve to the $E_7$ case, which requires delicate handling of degeneracies.
By combining the results of these two cases, we propose a ``local'' rule for the brane transitions.
\end{abstract}

\end{titlepage}

\tableofcontents

\section{Introduction}

It is a famous result by Hanany and Witten \cite{HW} that, in type IIB string theory, when an NS5-brane and a D5-brane placed at a different position in a line and perpendicular to each other (preserving supersymmetries) move across each other, a D3-brane stretching between them is generated.
As already explained in \cite{HW}, this can be interpreted as a conservation law of the 5-brane charges between both directions along the line.
The Hanany-Witten brane transition was later extended to brane systems with general types of 5-branes tilted relatively according to the types of the 5-branes.
It turns out to be a powerful tool in understanding dynamics of many supersymmetric gauge theories.

Indeed, exactly the same brane configurations appear in understanding the worldvolume theory of multiple M2-branes \cite{ABJM}.
It was known that, if we place $N$ D3-branes on a circle $S^1$ along with an NS5-brane and a $(1,k)$5-brane placed at a different position in the circle and tilted relatively by an angle depending on $k(>0)$, the brane configuration is described by the ${\cal N}=6$ Chern-Simons theory with gauge group $\text{U}(N)_k\times\text{U}(N)_{-k}$ (the subscripts $(k,-k)$ denoting the Chern-Simons levels) and two pairs of bifundamental matters $(N,\bar N)$ and $(\bar N,N)$. 
After taking the T-duality along the circle and lifting to M-theory, the system becomes that of $N$ multiple M2-branes placed on ${\mathbb C}^4/{\mathbb Z}_k$.
The system was further generalized by allowing rank differences of D3-branes between the two intervals of 5-branes \cite{HLLLP2,ABJ}.
Namely it was proposed that the ${\cal N}=6$ Chern-Simons theory with gauge group $\text{U}(N_1)_k\times\text{U}(N_2)_{-k}$ and two pairs of bifundamental matters describes the worldvolume theory of multiple M2-branes with $\min(N_1,N_2)$ M2-branes and $|N_2-N_1|$ fractional M2-branes on ${\mathbb C}^4/{\mathbb Z}_k$.

By applying the localization techniques \cite{P,KWY} the partition function on $S^3$ originally defined by the infinite-dimensional path integral reduces to a finite-dimensional matrix integral called the ABJM matrix model.
Using this matrix model we can reproduce the $N^{\frac{3}{2}}$ behavior of degrees of freedom \cite{DMP1} predicted from the gravity analysis \cite{KT}.
Besides, it was found that all of the perturbative corrections are summed up to an Airy function \cite{FHM}.

The discovery of the Airy function is just a starting point of the full explorations.
The integral expression for the Airy function suggests to regard the overall rank $N$ as a particle number and move to the grand canonical ensemble
\begin{align}
\Xi_k(z)=\sum_{N=0}^\infty z^NZ_k(N),
\end{align}
by introducing the dual fugacity $z$.
It was found \cite{MP} that the partition function of the ABJM matrix model can be rewritten into the partition function of a Fermi gas system, which is given by the Fredholm determinant
\begin{align}
\Xi_k(z)=\Det(1+z\widehat H^{-1}),
\label{HQP}
\end{align}
in the grand canonical ensemble with the spectral operator being $\widehat H=\widehat{\cal Q}\widehat{\cal P}$.
According to \cite{MP,MN1} we can translate the $(1,k)$5-brane and the NS5-brane respectively into
\begin{align}
\widehat{\cal Q}=\widehat Q^{\frac{1}{2}}+\widehat Q^{-\frac{1}{2}},\quad
\widehat{\cal P}=\widehat P^{\frac{1}{2}}+\widehat P^{-\frac{1}{2}},
\end{align}
which are given by exponentiating the canonical operators
\begin{align}
\widehat Q=e^{\widehat q},\quad
\widehat P=e^{\widehat p},\quad
[\widehat q,\widehat p]=i\hbar,
\end{align}
with the identification $\hbar=2\pi k$.
The spectral operator $\widehat H=(\widehat Q^{\frac{1}{2}}+\widehat Q^{-\frac{1}{2}})(\widehat P^{\frac{1}{2}}+\widehat P^{-\frac{1}{2}})$ is reminiscent of the defining equation for ${\mathbb P}^1\times{\mathbb P}^1$, which leads to the idea of quantum curves \cite{MiMo,ACDKV}.

By applying the WKB expansion to the Fermi gas formalism for the ABJM matrix model \eqref{HQP}, we can not only reproduce the Airy function easily but also proceed to study non-perturbative effects.
Namely, after the worldsheet instantons \cite{DMP1} and the membrane instantons \cite{DMP2} were detected from the 't Hooft expansion, the membrane instantons were systematically analyzed by the WKB expansion \cite{MP,CM}.
Besides, we can also utilize the Fermi gas formalism to find several exact values \cite{HMO1,PY} and obtain the full picture of the non-perturbative effects \cite{HMO2}.
Here, bound states of the two instantons were detected and found to be taken care of purely by the worldsheet instantons if we redefine the chemical potential $\log z$ suitably \cite{HMO3}.
Finally all of the non-perturbative effects are expressed by the free energy of topological strings along with the derivative of its refinement in the Nekrasov-Shatashvili limit on the background of local ${\mathbb P}^1\times{\mathbb P}^1$ \cite{HMMO}.

To summarize for now, on one hand the ABJM matrix model in the grand canonical ensemble is the Fredholm determinant with the spectral operator for ${\mathbb P}^1\times{\mathbb P}^1$.
On the other hand, the full non-perturbative expansions of the grand partition function of the ABJM matrix model are expressed by the free energy of topological strings on local ${\mathbb P}^1\times{\mathbb P}^1$.
Interestingly, by removing the role of the ABJM matrix model, it was proposed generally \cite{GHM1} that the Fredholm determinant with spectral operators for del Pezzo geometries is described by the free energy of topological strings on the same local geometries \cite{HKRS,GKMR,FMS,M}.

From the non-zero area of the phase space for the Fermi gas system, it is clear that the perturbative part of the grand potential $\log\Xi_k(z)$ is always a cubic polynomial of the chemical potential $\log z$, which directly implies the Airy function and the $N^{\frac{3}{2}}$ behavior in the canonical ensemble.
If we regard the Airy function as the characteristics of the multiple M2-brane systems, we may want to claim that the multiple M2-branes are described not by matrix models but by spectral operators or quantum curves \cite{M}.

Furthermore, as a bonus, compared with the original description by matrix models, spectral operators for del Pezzo geometries enjoy large symmetries of Weyl groups of exceptional groups.
For this reason, when we connect spectral operators with brane configurations directly, large symmetries of the spectral operators may imply unexpected symmetries for the brane configurations beyond the Hanany-Witten transitions.

These expectations were put on a firmer ground by studying a generalization of the ABJM matrix model.
Namely, instead of the original brane configurations leading to the ABJM matrix model, let us double the numbers of the two types of 5-branes, where the spectral operators apparently fall into the $D_5$ quantum curve.
Then, it was found that, if we place the same types of 5-branes next to each other (leading to the spectral operator $\widehat H=\widehat{\cal Q}^2\widehat{\cal P}^2$) \cite{MN3}, the instanton coefficients can be regarded as the BPS indices of the $D_5$ del Pezzo geometry \cite{KKV,HKP}.
Compared with it, the interpretation of the instanton effects is not as clear if the two types of 5-branes are placed alternatingly (leading to $\widehat H=\widehat{\cal Q}\widehat{\cal P}\widehat{\cal Q}\widehat{\cal P}$) \cite{HM}.

To study the spectral operator $\widehat H=\widehat{\cal Q}\widehat{\cal P}\widehat{\cal Q}\widehat{\cal P}$ in the same framework as $\widehat H=\widehat{\cal Q}^2\widehat{\cal P}^2$, we need to exchange 5-branes by applying the Hanany-Witten transitions.
As noted at the beginning, this is equivalent to introducing rank deformations.
In the rank deformations the BPS indices are split according to various combinations of the K\"ahler parameters \cite{MNN}.
This split of the BPS indices is later understood systematically as the decomposition of representations.
Namely, the BPS indices of the $D_5$ del Pezzo geometry form representations of $D_5$ and, if we assume a suitable subgroup for each rank deformation, the split of the BPS indices is understood as the decomposition of the representations into the subgroup \cite{MNY}.
Furthermore, after the Weyl group action on the quantum curve was identified, it was found \cite{KMN} that the assumed subgroup is nothing but the invariant subgroup of $D_5$ which preserves the quantum curve with the rank deformations.

After these findings, it was tantalizing to uncover the group structure of brane configurations for M2-branes by identifying the three-dimensional space of relative rank differences in the brane configurations with two NS5-branes and two $(1,k)$5-branes, ${\cal C}_\text{B}^{2,2}$, in the five-dimensional parameter space for the $D_5$ quantum curve, ${\cal C}_\text{P}^{D_5}$, (known as point configurations \cite{KNY}).
This question was investigated in \cite{KM} by resolving a previous confusion.
Since we are considering the Hanany-Witten transitions in a circle instead of in a line, the conservation law between both directions does not make sense.
Due to this reason, it was proposed in \cite{KM} to specify the interval which 5-branes do not move across in the Hanany-Witten transitions in identifying ${\cal C}_\text{B}^{2,2}$ in ${\cal C}_\text{P}^{D_5}$.
This concept is similar to fixing a local chart in manifolds or fixing a local reference frame in mechanical systems.

After embedding ${\cal C}_\text{B}^{2,2}$ in ${\cal C}_\text{P}^{D_5}$, we find that the subgroup of $W(D_5)$ (the Weyl group of $D_5$) transforming among the three-dimensional subspace ${\cal C}_\text{B}^{2,2}$ is $W(B_3)$.
It is then natural to ask whether all the symmetries in $W(B_3)$ are known from the established Hanany-Witten transitions.
After separating the transformations coming from the Hanany-Witten transitions, it was found that there are brane transitions mapping the brane configuration $\langle N_1\bullet N_2\circ N'_3\bullet N_4\circ\rangle$ into $\langle N_1\bullet N'_3\circ N_2\bullet N_4\circ\rangle$ or $\langle N_1\bullet N_2\circ N_4\bullet N'_3\circ\rangle$, whose interpretation was not known previously.
Here $\bullet$ and $\circ$ denote the NS5-brane and the $(1,k)$5-brane respectively and the integer is the number of D3-branes in each interval, while the bracket $\langle\cdots\rangle$ denotes the reference interval which the 5-branes do not move across in the Hanany-Witten transitions.
Note that, from the analysis so far, we can only claim the brane transitions within the whole set of the brane configurations on the circle $\langle N_1\bullet N_2\circ N'_3\bullet N_4\circ\rangle$.
It is of course interesting to understand these brane transitions ``locally'' (without referring to the whole configurations, as the Hanany-Witten transitions) and explore the full structure of them, though it is difficult (or dangerous) to study simply from this example.

In this paper, motivated by understanding the new brane transitions, we proceed to the space of more complicated brane configurations ${\cal C}_\text{B}^{2,4}$ with two NS5-branes and four $(1,k)$5-branes, which are realized by the $E_7$ quantum curve.
We stress that the analysis, however, is not a simple generalization of the previous $D_5$ case due to degeneracies of the $E_7$ curve.
For this reason, not all of the brane configurations with the same numbers of 5-branes are realized by the $E_7$ quantum curve, but only a subclass of them are.
We identify the subspace of brane configurations in ${\cal C}_\text{B}^{2,4}$ which can be realized by the $E_7$ quantum curve, and study the realization in the space of point configurations ${\cal C}_\text{P}^{E_7}$ explicitly.
By combining the results of both the $D_5$ and $E_7$ cases, we can propose a simple rule for the brane transitions within brane configurations.
With this proposal, the inclusion relation ${\cal C}_\text{P}^{E_7}\subset{\cal C}_\text{B}^{2,4}$ may give us a clue to understand the brane transitions beyond the $E_7$ quantum curve in ${\cal C}_\text{P}^{E_7}$.

The results are summarized as follows.
First, we find that the four-dimensional subspace of the five-dimensional space of rank deformations in brane configurations ${\cal C}_\text{B}^{2,4}$ is realized in the seven-dimensional space of point configurations ${\cal C}_\text{P}^{E_7}$.
The subgroup of $W(E_7)$ transforms among the four-dimensional subspace of brane configurations ${\cal C}_\text{B}^{2,4}\cap{\cal C}_\text{P}^{E_7}$ is $W(F_4)$.
In clarifying the correspondence between brane configurations and point configurations, we also provide several intuitive understandings for previous observations.
For example, we can explain with the ${\mathbb Z}_2$ folding of Dynkin diagrams the invariant subgroups $W(B_3)$ and $W(F_4)$ remaining in the subspaces realized by the quantum curves.

After separating the transformations coming from the Hanany-Witten transitions, again we can specify new brane transitions which are not explained from the known transitions.
By comparing the new brane transitions between the $D_5$ case studied previously in \cite{KM} and the $E_7$ case studied in this paper, we tentatively propose a ``local'' rule for the brane transitions
\begin{align}
\cdots\circ N\bullet N'\circ\cdots=\cdots\circ N'\bullet N\circ\cdots,\quad
\cdots\bullet N\circ N'\bullet\cdots=\cdots\bullet N'\circ N\bullet\cdots,
\label{localintro}
\end{align}
where all of the remaining parts denoted by $\dots$ are equal on both sides.
Physically it is surprising for us to observe these brane transitions.
Since the ordering of D3-branes may change the brane dynamics drastically, the brane transitions \eqref{localintro} exchanging the neighboring numbers of D3-branes may hold only restrictively.
It is an important future direction to specify the situations when the brane transitions are valid and clarify whether or not it is accidental only for curves of genus one since their origins are the exceptional Weyl groups characteristic for these curves.

The organization of the paper is as follows.
After recapitulating the brane transitions found for the $D_5$ quantum curve in section \ref{d5}, we embark on our study of the brane transitions for the $E_7$ quantum curve in section \ref{e7}.
By comparing the new brane transitions for these two cases, in the subsequent section we propose the local rule for the brane transitions \eqref{localintro} and provide some checks and discussions for it.
Finally we conclude with some discussions on future directions.

\section{Review: $D_5$ curve}\label{d5}

In this section, we recapitulate how the new brane transitions were found from the $D_5$ quantum curve.
After reviewing the $D_5$ quantum curve and the brane configurations, we explain how the space of brane configurations ${\cal C}_\text{B}^{2,2}$ with two NS5-branes and two $(1,k)$5-branes is embedded in the space of point configurations ${\cal C}_\text{P}^{D_5}$ for the $D_5$ curve.

\subsection{Quantum curve and Weyl group}

\begin{figure}[!t]
\centering\includegraphics[scale=0.5,angle=-90]{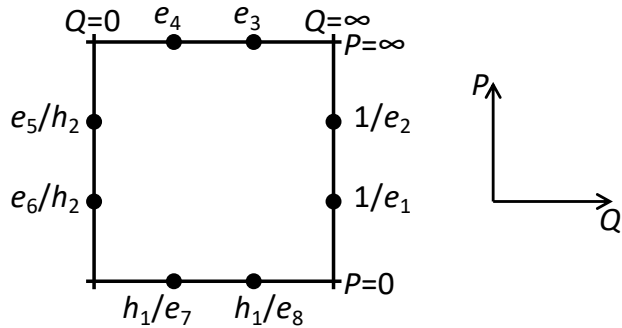}
\caption{Asymptotic values for the $D_5$ curve.}
\label{d5asympt}
\end{figure}

We review the $D_5$ quantum curve and its symmetry in this subsection.
It is well-known that the classical $D_5$ curve is generated by the nine terms consisting of $Q^mP^n$ $(m,n=-1,0,+1)$.
In \cite{KMN} the variables $Q$ and $P$ are lifted to quantum-mechanical operators
\begin{align}
\widehat Q=e^{\widehat q},\quad\widehat P=e^{\widehat p},\quad[\widehat q,\widehat p]=i\hbar,
\label{ccr}
\end{align}
satisfying
\begin{align}
\widehat P^\beta\widehat Q^\alpha=q^{-\alpha\beta}\widehat Q^\alpha\widehat P^\beta,\quad q=e^{i\hbar},
\label{cr}
\end{align}
with the Planck constant $\hbar$ identified with the Chern-Simons level $k$ by
\begin{align}
\hbar=2\pi k.
\end{align}
Then, the quantum-mechanical spectral operator for the $D_5$ curve is given by
\begin{align}
&\widehat H/\alpha
=q^{-\frac{1}{2}}\widehat Q^{-1}(\widehat Q+q^{\frac{1}{2}}e_3)(\widehat Q+q^{\frac{1}{2}}e_4)\widehat P
+(e_1^{-1}+e_2^{-1})\widehat Q+E/\alpha
\nonumber\\&\quad
+e_3e_4h_2^{-1}(e_5+e_6)\widehat Q^{-1}
+q^{\frac{1}{2}}(e_1e_2)^{-1}\widehat Q^{-1}(\widehat Q+q^{-\frac{1}{2}}h_1e_7^{-1})(\widehat Q+q^{-\frac{1}{2}}h_1e_8^{-1})\widehat P^{-1},
\label{d5curve}
\end{align}
with the ten parameters satisfying the constraint $(h_1h_2)^2=e_1e_2e_3e_4e_5e_6e_7e_8$.
The appearance of the parameter $q$ in \eqref{d5curve} is due to the introduction of the Weyl order.
The asymptotic values for the classical cousin of the curve are depicted in figure \ref{d5asympt}.
The $D_5$ quantum curve enjoys trivial symmetries of exchanging the asymptotic values
\begin{align}
s_1:h_1e_7^{-1}\leftrightarrow h_1e_8^{-1},\quad s_2:e_3\leftrightarrow e_4,\quad
s_5:e_1^{-1}\leftrightarrow e_2^{-1},\quad s_0:h_2^{-1}e_5\leftrightarrow h_2^{-1}e_6.
\end{align}
Besides, due to non-trivial similarity transformations, we can also exchange the asymptotic values on the opposite sides
\begin{align}
s_3:e_3\leftrightarrow h_1e_7^{-1},\quad s_4:e_1^{-1}\leftrightarrow h_2^{-1}e_5.
\end{align}

\begin{figure}[!t]
\centering\includegraphics[scale=0.5,angle=-90]{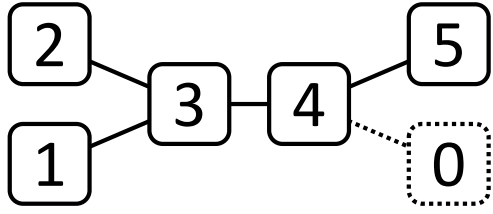}
\caption{Dynkin diagram of the $D_5$ algebra.}
\label{d5alg}
\end{figure}

Since we have used ten parameters subject to one constraint to parametrize eight asymptotic values subject to the same constraint, there should be two degrees of freedom to be fixed.
Also note that the spectral operator $\widehat H$ is invariant under similarity transformations, which contain
\begin{align}
\widehat Q\to A\widehat Q,\quad
\widehat P\to B\widehat P,
\label{AB}
\end{align}
with arbitrary coefficients $A,B$.
This gives another two degrees of freedom to be fixed.
If we fix the four degrees of freedom and solve the constraint by
\begin{align}
e_2=e_4=e_6=e_8=1,\quad
e_7=(h_1h_2)^2/(e_1e_3e_5),
\end{align}
the $D_5$ curve is characterized by the remaining parameters in
\begin{align}
{\cal C}_\text{P}^{D_5}=\{(h_1,h_2,e_1,e_3,e_5)\}.
\end{align}
Then, the similarity transformations are given unambiguously by
\begin{align}
s_1&:(h_1,h_2,e_1,e_3,e_5)\mapsto((h_1h_2^2)^{-1}e_1e_3e_5,h_2,e_1,e_3,e_5),\nonumber\\
s_2&:(h_1,h_2,e_1,e_3,e_5)\mapsto(h_1e_3^{-1},h_2,e_1,e_3^{-1},e_5),\nonumber\\
s_3&:(h_1,h_2,e_1,e_3,e_5)\mapsto(h_1,(h_1h_2)^{-1}e_1e_5,e_1,(h_1h_2^2)^{-1}e_1e_3e_5,e_5),\nonumber\\
s_4&:(h_1,h_2,e_1,e_3,e_5)\mapsto(h_1h_2(e_1e_5)^{-1},h_2,h_2e_5^{-1},e_3,h_2e_1^{-1}),\nonumber\\
s_5&:(h_1,h_2,e_1,e_3,e_5)\mapsto(h_1,h_2e_1^{-1},e_1^{-1},e_3,e_5),\nonumber\\
s_0&:(h_1,h_2,e_1,e_3,e_5)\mapsto(h_1,h_2e_5^{-1},e_1,e_3,e_5^{-1}),
\end{align}
on these parameters.
These transformations form the $D_5$ Weyl group whose Dynkin diagram (along with the numberings corresponding to the transformations) is given in figure \ref{d5alg}.

\subsection{Brane configurations}

So far we have explained the $D_5$ quantum curve and its symmetry.
Now let us turn to the corresponding brane configurations.

\begin{figure}[!t]
\centering\includegraphics[scale=0.5,angle=-90]{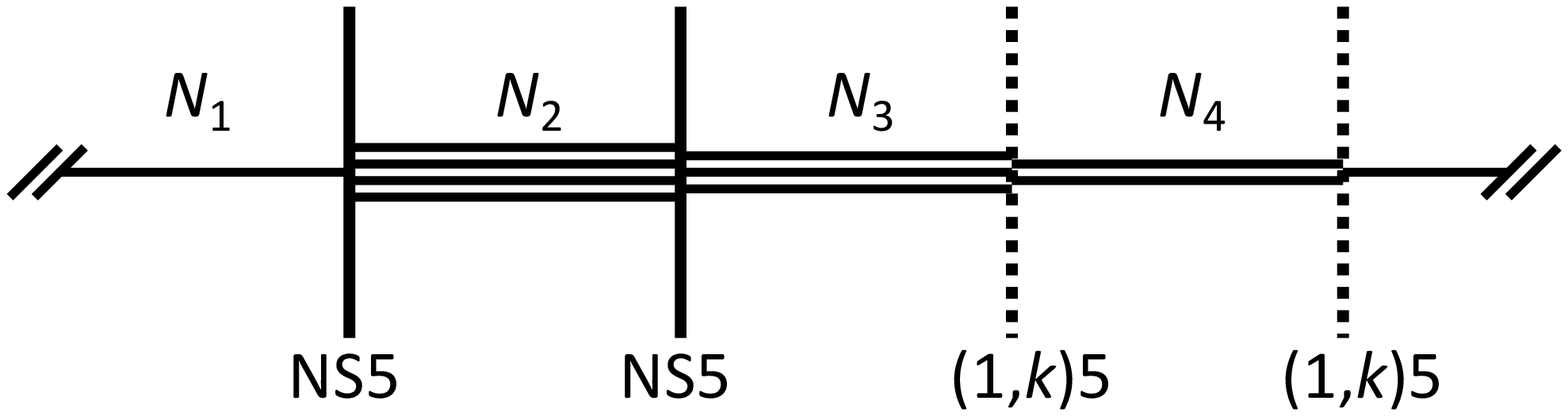}
\caption{Brane configuration denoted by $\langle N_1\stackrel{2}{\bullet}N_2\stackrel{1}{\bullet}N_3\stackrel{3}{\circ}N_4\stackrel{4}{\circ}\rangle$.}
\label{bc}
\end{figure}

In the setup for multiple M2-branes, we compactify one dimension of $N$ D3-branes on a circle $S^1$ and place NS5-branes and $(1,k)$5-branes perpendicular to the original D3-branes and tilted relatively by an angle determined by $k(>0)$ to preserve the supersymmetries.
For the case of the $D_5$ quantum curve the corresponding brane configurations contain two NS5-branes and two $(1,k)$5-branes where we label the NS5-branes by $1,2$ and the $(1,k)$5-branes by $3,4$ respectively.
We express the brane configuration by
\begin{align}
&\langle N_1\stackrel{2}{\bullet}N_2\stackrel{1}{\bullet}N_3\stackrel{3}{\circ}N_4\stackrel{4}{\circ}\rangle\nonumber\\
&\quad=\langle N+M_2+M_3\stackrel{2}{\bullet}N+M_1+2M_3\stackrel{1}{\bullet}N+2M_1+M_2+M_3\stackrel{3}{\circ}N+M_1\stackrel{4}{\circ}\rangle,
\label{NM1M2M3}
\end{align}
where $\bullet$ and $\circ$ denote the NS5-brane and the $(1,k)$5-brane respectively and the integer is the number of D3-branes in each interval (see figure \ref{bc} for the brane picture for $\langle N_1\stackrel{2}{\bullet}N_2\stackrel{1}{\bullet}N_3\stackrel{3}{\circ}N_4\stackrel{4}{\circ}\rangle$).
Note that, as explained in \cite{KM}, in discussing the brane configurations we need to fix the reference interval the 5-branes do not move across in the Hanany-Witten transitions, which we denote by the bracket $\langle\cdots\rangle$.
By solving inversely, we find that the relative ranks are given by
\begin{align}
M_1=\frac{-N_1+N_3}{2},\quad
M_2=\frac{N_1-N_2+N_3-N_4}{2},\quad
M_3=\frac{N_2-N_4}{2}.
\label{M1M2M3}
\end{align}
Hence, the brane configurations are parametrized by these relative ranks,
\begin{align}
{\cal C}_\text{B}^{2,2}=\{(M_1,M_2,M_3)\}.
\end{align}
For simplicity, we assume the level $k$ to be large enough and do not consider the situations leading to duality cascades \cite{KS} or nonsupersymmetric configurations.
(See \cite{HK} for recent arguments.)

\subsection{Brane transitions}

In this subsection, we explain how new brane transitions are found by connecting the point configurations for the $D_5$ curve ${\cal C}_\text{P}^{D_5}$ and the rank differences in the brane configurations ${\cal C}_\text{B}^{2,2}$ explained in the previous two subsections.

In the Fermi gas formalism for the generalizations of the ABJM matrix model without rank deformations, it was found that the $(1,k)$5-brane and the NS5-brane are translated respectively into the canonical operators
\begin{align}
\widehat{\cal Q}=\widehat Q^{\frac{1}{2}}+\widehat Q^{-\frac{1}{2}},\quad
\widehat{\cal P}=\widehat P^{\frac{1}{2}}+\widehat P^{-\frac{1}{2}},
\end{align}
satisfying the commutation relation \eqref{cr} and, for the brane configurations (with the common overall rank $N$ abbreviated)
\begin{align}
\langle\underbrace{\bullet\bullet\cdots\bullet}_{p_1}\underbrace{\circ\circ\cdots\circ}_{q_1}
\underbrace{\bullet\bullet\cdots\bullet}_{p_2}\underbrace{\circ\circ\cdots\circ}_{q_2}\cdots\rangle,
\end{align}
the spectral operator $\widehat H$ appearing in the Fredholm determinant $\Xi_k(z)=\Det(1+z\widehat H^{-1})$ for the grand partition function $\Xi_k(z)=\sum_{N=0}^\infty z^NZ_k(N)$ is given by
\begin{align}
\widehat H=\cdots\widehat{\cal Q}^{q_2}\widehat{\cal P}^{p_2}\widehat{\cal Q}^{q_1}\widehat{\cal P}^{p_1},
\end{align}
(as reviewed carefully in \cite{KM}).

Using simply the relation without rank deformations, we can identify the relation for rank deformations and embed the three-dimensional space of brane configurations ${\cal C}_\text{B}^{2,2}$ into the five-dimensional space of point configurations ${\cal C}_\text{P}^{D_5}$.
The key tool is the Hanany-Witten brane transitions.
Namely, on one hand, for each quantum curve, we can identify the parameters of the point configuration in ${\cal C}_\text{P}^{D_5}$ by comparing with the general expressions \eqref{d5curve}.
On the other hand, for the corresponding brane configuration, we can apply the Hanany-Witten transitions
\begin{align}
&\cdots K\circ L\circ M\cdots=\cdots K\circ K+M-L\circ M\cdots,\nonumber\\
&\cdots K\bullet L\bullet M\cdots=\cdots K\bullet K+M-L\bullet M\cdots,\nonumber\\
&\cdots K\circ L\bullet M\cdots=\cdots K\bullet K+M-L+k\circ M\cdots,\nonumber\\
&\cdots K\bullet L\circ M\cdots=\cdots K\circ K+M-L+k\bullet M\cdots,
\label{HW}
\end{align}
to bring the 5-branes into the standard order $\langle N_1\stackrel{2}{\bullet}N_2\stackrel{1}{\bullet}N_3\stackrel{3}{\circ}N_4\stackrel{4}{\circ}\rangle$ and obtain the relative ranks \eqref{M1M2M3} in ${\cal C}_\text{B}^{2,2}$.
As was noted in \cite{KM}, since, on the brane configuration side, it is important to fix the reference interval where exchanges of 5-branes are forbidden, correspondingly on the point configuration side, we need to avoid general similarity transformations and allow only ``small'' similarity transformations given in \eqref{AB}.

The first observation is that all the point configurations obtained from the brane configurations ${\cal C}_\text{B}^{2,2}$ are located within a three-dimensional subspace of the original five-dimensional space ${\cal C}_\text{P}^{D_5}$ which can be labelled by
\begin{align}
(h_1,h_2,e_1,e_3,e_5)=\biggl(\frac{m_2m_3}{m_1},\frac{m_1m_3}{m_2},\frac{m_3}{m_2},m_2m_3,\frac{m_3}{m_2}\biggr).
\label{hheee}
\end{align}
We will come back to the intuitive explanation for this embedding in the next section.
In relating the rank deformations $(M_1,M_2,M_3)$ to the parameters of the $D_5$ curve $(m_1,m_2,m_3)$, we find
\begin{align}
(m_1,m_2,m_3)=(e^{2\pi i(M_1-k)},e^{2\pi iM_2},e^{2\pi iM_3}).
\label{hef}
\end{align}

\begin{figure}[!t]
\centering\includegraphics[scale=0.5,angle=-90]{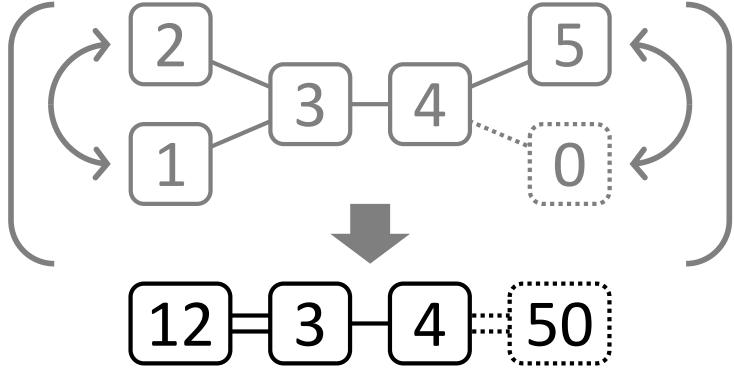}
\caption{Dynkin diagram of the $B_3$ algebra (denoting the symmetries for the three-dimensional subspace of brane configurations $C_\text{B}^{2,2}$).
This can be regarded as the ${\mathbb Z}_2$ folding of the $D_5$ Dynkin diagram.}
\label{b3alg}
\end{figure}

Among others, it is interesting to identify the trivial transformations which leave each point of ${\cal C}_\text{B}^{2,2}$ invariant and, after modding out these trivial transformations, the transformations which keep the subspace ${\cal C}_\text{B}^{2,2}$ invariant.
In \cite{KM} it was found that the former forms $W((A_1)^2)$ generated by $s_1s_3s_4s_5s_4s_3s_1$ and $s_2s_3s_4s_5s_4s_3s_2$ while the latter forms $W(B_3)$ generated by $s_1s_2$, $s_3$ and $s_4$ (as well as $s_5s_0$) (see figure \ref{b3alg} for the Dynkin diagram).
The ranks of two algebras seem consistent with the dimensions of the perpendicular and parallel directions for the three-dimensional subspace of brane configurations ${\cal C}_\text{B}^{2,2}$ in the five-dimensional space of point configurations ${\cal C}_\text{P}^{D_5}$.
Note that the $B_3$ Dynkin diagram can be regarded as the ${\mathbb Z}_2$ folding of the $D_5$ Dynkin diagram.
The interpretations of this folding will be explained in the next section.
Among $W(B_3)$, $s_1s_2$ and $s_5s_0$ are identified respectively as the Hanany-Witten transition exchanging the two $(1,k)$5-branes and that exchanging the two NS5-branes.
It remains to understand the transformations $s_3$ and $s_4$, which are given by
\begin{align}
s_3&:\langle N_1\stackrel{2}{\bullet}N_2\stackrel{1}{\bullet}N_3\stackrel{3}{\circ}N_4\stackrel{4}{\circ}\rangle\mapsto
\langle N_1\stackrel{2}{\bullet}N_2-N_3+N_4+k\stackrel{1}{\bullet}-N_3+2N_4+2k\stackrel{3}{\circ}N_4\stackrel{4}{\circ}\rangle,\nonumber\\
s_4&:\langle N_1\stackrel{2}{\bullet}N_2\stackrel{1}{\bullet}N_3\stackrel{3}{\circ}N_4\stackrel{4}{\circ}\rangle\mapsto
\langle N_1\stackrel{2}{\bullet}N_2\stackrel{1}{\bullet}2N_2-N_3+2k\stackrel{3}{\circ}N_2-N_3+N_4+k\stackrel{4}{\circ}\rangle,
\label{d5N}
\end{align}
in terms of brane configurations.
By defining $N'_3=N_2-N_3+N_4+k$ we can rewrite the brane transitions as
\begin{align}
s_3&:\langle N_1\stackrel{2}{\bullet}N_2\stackrel{3}{\circ}N'_3\stackrel{1}{\bullet}N_4\stackrel{4}{\circ}\rangle\mapsto
\langle N_1\stackrel{2}{\bullet}N'_3\stackrel{3}{\circ}N_2\stackrel{1}{\bullet}N_4\stackrel{4}{\circ}\rangle,\nonumber\\
s_4&:\langle N_1\stackrel{2}{\bullet}N_2\stackrel{3}{\circ}N'_3\stackrel{1}{\bullet}N_4\stackrel{4}{\circ}\rangle\mapsto
\langle N_1\stackrel{2}{\bullet}N_2\stackrel{3}{\circ}N_4\stackrel{1}{\bullet}N'_3\stackrel{4}{\circ}\rangle.
\label{d5s3s4}
\end{align}
Here the $s_3$ transformation exchanges the numbers of D3-branes on two sides of the $(1,k)$5-brane $\stackrel{3}{\circ}$ surrounded by two NS5-branes $\bullet$, while the $s_4$ transformation exchanges those on two sides of the NS5-branes $\stackrel{1}{\bullet}$ surrounded by two $(1,k)$5-branes $\circ$.

The Hanany-Witten transitions are well-known in brane configurations.
By fixing the reference interval, we utilize the Hanany-Witten transitions to bring 5-branes into the standard order and relate relative ranks of brane configurations to parameters of point configurations.
After the identification, since the space of point configurations ${\cal C}_\text{P}^{D_5}$ enjoy the symmetry of $W(D_5)$, the space of brane configurations ${\cal C}_\text{B}^{2,2}$ partially inherits the Weyl group $W(B_3)$.
After separating the well-known Hanany-Witten transitions we can identify the new brane transitions \eqref{d5s3s4} generated from the transformations $s_3$ and $s_4$.
It is desirable to understand these brane transitions more systematically. 
Simply from these transformations, however, it is dangerous to guess general rules for the brane transitions.
In the next section we shall repeat the same analysis for the $E_7$ quantum curve to unveil the structure.

\section{$E_7$ curve}\label{e7}

In the previous section, we have compared the brane configurations with two NS5-branes and two $(1,k)$5-branes with the point configurations for the $D_5$ quantum curve and identified the unknown brane transitions.
It is desirable to study these brane transitions more systematically, though it is difficult to understand the full structure simply from this example.
In this section, we shall repeat the same analysis for the $E_7$ curve.
Here, unlike $E_6$ or $E_8$, the $E_7$ curve has a rectangular toric realization with its center located at the origin and is suitable for our current analysis of brane configurations.
After recapitulating the $E_7$ quantum curve and clarifying the corresponding brane configurations with careful handling of degeneracies of the $E_7$ curve, we again embed the space of brane configurations in the space of point configurations and identify the unknown symmetries beyond the Hanany-Witten transitions.
With the analysis in this section, we can present intuitive understandings in how brane configurations are embedded in point configurations.
Besides, our results here enable discussions on systematical understanding of the new brane transitions in the next section.

\subsection{Quantum curve and Weyl group}

\begin{figure}[!t]
\centering\includegraphics[scale=0.5,angle=-90]{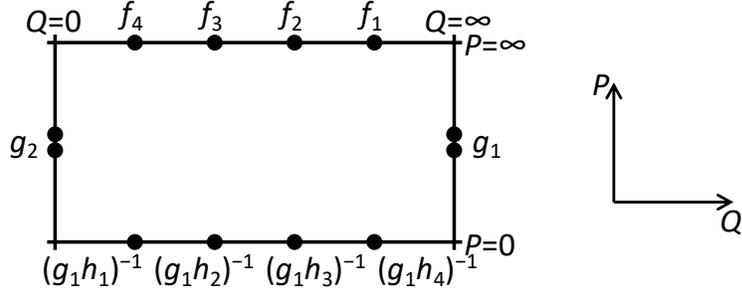}
\caption{Asymptotic values for the $E_7$ curve.
The classical doubly-degenerate asymptotic values $g_1$ and $g_2$ are resolved quantum-mechanically.}
\label{e7asympt}
\end{figure}

We shall first review the $E_7$ quantum curve in this subsection.
The explicit expression of the $E_7$ quantum curve was given previously in \cite{KMN,M}.
After introducing ten parameters $(f_1,f_2,f_3,f_4,g_1,g_2,h_1,h_2,h_3,h_4)$ subject to the constraint $f_1f_2f_3f_4(g_1g_2)^2h_1h_2h_3h_4=1$, we can write down the $E_7$ quantum curve as
\begin{align}
&\widehat H/\alpha=q^{-1}\widehat Q^{-2}\prod_{i=1}^4(\widehat Q+q^{\frac{1}{2}}f_i)\widehat P
+[2]_qg_1\widehat Q^2+(F_1g_1+F_4G_2^2H_3)\widehat Q+E/\alpha
\nonumber\\&\quad
+(F_3g_2+F_4G_2H_1)\widehat Q^{-1}+[2]_qF_4g_2\widehat Q^{-2}
+qg_1^2\widehat Q^{-2}\prod_{i=1}^4(\widehat Q+q^{-\frac{1}{2}}(g_1h_i)^{-1})\widehat P^{-1},
\label{e7curve}
\end{align}
with the $q$-integer $[2]_q=q^{\frac{1}{2}}+q^{-\frac{1}{2}}$ and the coefficients
\begin{align}
\sum_{n=0}^4F_nx^n=\prod_{i=1}^4(1+f_ix),\quad
\sum_{n=0}^2G_nx^n=\prod_{i=1}^2(1+g_ix),\quad
\sum_{n=0}^4H_nx^n=\prod_{i=1}^4(1+h_ix),
\end{align}
subject to the constraint
\begin{align}
F_4G_2^2H_4=1.
\label{e7constraint}
\end{align}
The asymptotic values for the $E_7$ curve are depicted in figure \ref{e7asympt}.
Here the asymptotic values $g_1$ and $g_2$ are doubly degenerate, which causes nontrivial restrictions in brane configurations, as studied in the next subsection.
We can choose the condition
\begin{align}
f_4=g_2=1,\quad
h_4=(f_1f_2f_3g_1^2h_1h_2h_3)^{-1},
\label{e7gauge}
\end{align}
by using the ``small'' similarity transformation \eqref{AB} and solving the constraint \eqref{e7constraint}.
Then, it is not difficult to see that the $E_7$ curve is characterized by seven parameters in
\begin{align}
{\cal C}_\text{P}^{E_7}=\{(f_1,f_2,f_3,g_1,h_1,h_2,h_3)\}.
\end{align}

\begin{figure}[!t]
\centering\includegraphics[scale=0.5,angle=-90]{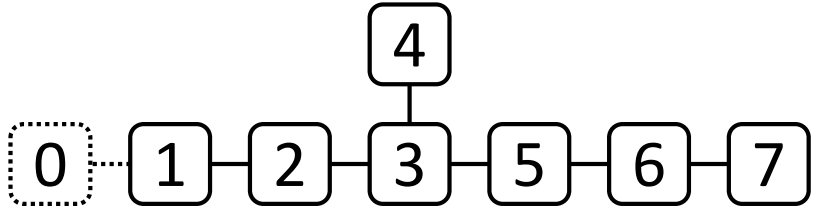}
\caption{Dynkin diagram of the $E_7$ Lie algebra.}
\label{e7alg}
\end{figure}

The $E_7$ quantum curve is invariant under similarity transformations isomorphic to the $E_7$ Weyl group, where after fixing the condition \eqref{e7gauge} the generators are given unambiguously by
\begin{align}
s_0&:(f_1,f_2,f_3,g_1,h_1,h_2,h_3)\mapsto(f_3^{-1}f_1,f_3^{-1}f_2,f_3^{-1},g_1,f_3h_1,f_3h_2,f_3h_3),\nonumber\\
s_1&:(f_1,f_2,f_3,g_1,h_1,h_2,h_3)\mapsto(f_1,f_3,f_2,g_1,h_1,h_2,h_3),\nonumber\\
s_2&:(f_1,f_2,f_3,g_1,h_1,h_2,h_3)\mapsto(f_2,f_1,f_3,g_1,h_1,h_2,h_3),\nonumber\\
s_3&:(f_1,f_2,f_3,g_1,h_1,h_2,h_3)\mapsto((g_1h_1)^{-1},f_2,f_3,(f_1h_1)^{-1},h_1,(f_1g_1h_1)h_2,(f_1g_1h_1)h_3),\nonumber\\
s_4&:(f_1,f_2,f_3,g_1,h_1,h_2,h_3)\mapsto(f_1,f_2,f_3,g_1^{-1},g_1h_1,g_1h_2,g_1h_3),\nonumber\\
s_5&:(f_1,f_2,f_3,g_1,h_1,h_2,h_3)\mapsto(f_1,f_2,f_3,g_1,h_2,h_1,h_3),\nonumber\\
s_6&:(f_1,f_2,f_3,g_1,h_1,h_2,h_3)\mapsto(f_1,f_2,f_3,g_1,h_1,h_3,h_2),\nonumber\\
s_7&:(f_1,f_2,f_3,g_1,h_1,h_2,h_3)\mapsto(f_1,f_2,f_3,g_1,h_1,h_2,(f_1f_2f_3g_1^2h_1h_2h_3)^{-1}).
\end{align}
See figure \ref{e7alg} for the $E_7$ Dynkin diagram along with the numberings.

\subsection{Restrictions for brane configurations}

After reviewing the $E_7$ quantum curve in the previous subsection, let us turn to the study of brane configurations.
The corresponding brane configurations contain two NS5-branes and four $(1,k)$5-branes and we denote the five-dimensional space of relative ranks in brane configurations as ${\cal C}_\text{B}^{2,4}$.
Note that unlike the $D_5$ case, not all the brane configurations with the same numbers of 5-branes without rank deformations fall into the $E_7$ quantum curve, but only a subclass of them are.
The degeneracies of the $E_7$ curve impose nontrivial restrictions for the brane configurations, which we clarify in this subsection.

In fact, only $\widehat{\cal Q}^2\widehat{\cal P}\widehat{\cal Q}^2\widehat{\cal P}$, $\widehat{\cal Q}\widehat{\cal P}\widehat{\cal Q}^2\widehat{\cal P}\widehat{\cal Q}$ or $\widehat{\cal P}\widehat{\cal Q}^2\widehat{\cal P}\widehat{\cal Q}^2$ are the $E_7$ quantum curve.
Due to the commutation relations
\begin{align}
&\widehat{\cal P}\widehat Q^{\frac{n}{2}}
=\widehat Q^{\frac{n}{2}}(q^{-\frac{n}{4}}\widehat P^{\frac{1}{2}}+q^{\frac{n}{4}}\widehat P^{-\frac{1}{2}}),\quad
\widehat{\cal P}\widehat Q^{-\frac{n}{2}}
=\widehat Q^{-\frac{n}{2}}(q^{\frac{n}{4}}\widehat P^{\frac{1}{2}}+q^{-\frac{n}{4}}\widehat P^{-\frac{1}{2}}),\nonumber\\
&\widehat P^{\frac{n}{2}}\widehat{\cal Q}
=(q^{-\frac{n}{4}}\widehat Q^{\frac{1}{2}}+q^{\frac{n}{4}}\widehat Q^{-\frac{1}{2}})\widehat P^{\frac{n}{2}},\quad
\widehat P^{-\frac{n}{2}}\widehat{\cal Q}
=(q^{\frac{n}{4}}\widehat Q^{\frac{1}{2}}+q^{-\frac{n}{4}}\widehat Q^{-\frac{1}{2}})\widehat P^{-\frac{n}{2}},
\label{calcr}
\end{align}
the asymptotic values of $\widehat P$ for large $\widehat Q$ or small $\widehat Q$ depend on the number of $\widehat{\cal Q}=\widehat Q^{\frac{1}{2}}+\widehat Q^{-\frac{1}{2}}$ located right to it, where each $\widehat{\cal Q}$ changes the power of the asymptotic values of $\widehat P$ by $q^{\pm\frac{1}{2}}$.
Since the $E_7$ quantum curve \eqref{e7curve} reads 
\begin{align}
&\widehat H/\alpha=q^{-1}\widehat Q^2(\widehat P+q^{\frac{3}{2}}g_1)(\widehat P+q^{\frac{1}{2}}g_1)\widehat P^{-1}
+\cdots+qF_4\widehat Q^{-2}(\widehat P+q^{-\frac{3}{2}}g_2)(\widehat P+q^{-\frac{1}{2}}g_2)\widehat P^{-1},
\end{align}
by collecting terms of the same order in $\widehat Q$, it is clear from the ratios $(q^{\frac{3}{2}}g_1)/(q^{\frac{1}{2}}g_1)=(q^{\frac{1}{2}})^2$ and $(q^{-\frac{3}{2}}g_2)/(q^{-\frac{1}{2}}g_2)=(q^{-\frac{1}{2}})^2$ that the brane configurations fall into the $E_7$ quantum curve if and only if there are exactly two $(1,k)$5-branes between the two NS5-branes.
Hence, we have the brane configurations $\langle\bullet\circ\circ\bullet\circ\,\circ\rangle$, $\langle\circ\bullet\circ\circ\bullet\,\circ\rangle$ and $\langle\circ\circ\bullet\circ\circ\,\bullet\rangle$, which correspond to the $E_7$ quantum curve $\widehat{\cal Q}^2\widehat{\cal P}\widehat{\cal Q}^2\widehat{\cal P}$, $\widehat{\cal Q}\widehat{\cal P}\widehat{\cal Q}^2\widehat{\cal P}\widehat{\cal Q}$ and $\widehat{\cal P}\widehat{\cal Q}^2\widehat{\cal P}\widehat{\cal Q}^2$.

\begin{table}[t!]
\begin{center}
\begin{tabular}{l||l|l|l|l}
type&$g_1$&$\{f_1,f_2,f_3,f_4\}$&$g_2$&$\{(g_1h_1)^{-1},(g_1h_2)^{-1},(g_1h_3)^{-1},(g_1h_4)^{-1}\}$\\\hline\hline
$\widehat{\cal Q}^2\widehat{\cal P}\widehat{\cal Q}^2\widehat{\cal P}$&$q^{-\frac{1}{2}}$&$\{1,1,q^{-\frac{1}{2}},q^{-\frac{1}{2}}\}$&$q^{\frac{1}{2}}$&$\{1,1,q^{\frac{1}{2}},q^{\frac{1}{2}}\}$\\
$\widehat{\cal Q}\widehat{\cal P}\widehat{\cal Q}^2\widehat{\cal P}\widehat{\cal Q}$&$1$&$\{q^{\frac{1}{2}},1,1,q^{-\frac{1}{2}}\}$&$1$&$\{q^{-\frac{1}{2}},1,1,q^{\frac{1}{2}}\}$\\
$\widehat{\cal P}\widehat{\cal Q}^2\widehat{\cal P}\widehat{\cal Q}^2$&$q^{\frac{1}{2}}$&$\{q^{\frac{1}{2}},q^{\frac{1}{2}},1,1\}$&$q^{-\frac{1}{2}}$&$\{q^{-\frac{1}{2}},q^{-\frac{1}{2}},1,1\}$\\
\end{tabular}
\end{center}
\caption{Asymptotic values for the $E_7$ quantum curve corresponding to the brane configurations.}
\label{e7asymptotic}
\end{table}

For these cases we can read off the asymptotic values as in table \ref{e7asymptotic}.
For example, we obtain the asymptotic values for $\widehat{\cal Q}^2\widehat{\cal P}\widehat{\cal Q}^2\widehat{\cal P}$ in the table from the expansion
\begin{align}
\widehat{\cal Q}^2\widehat{\cal P}\widehat{\cal Q}^2\widehat{\cal P}/q^{\frac{1}{2}}
=q^{-1}\widehat Q^{-2}(\widehat Q+1)^2(\widehat Q+q^{\frac{1}{2}})^2(\widehat P+1)
+\widehat Q^{-2}(\widehat Q+1)^2(\widehat Q+q^{-\frac{1}{2}})^2(1+\widehat P^{-1}).
\end{align}
by comparing with \eqref{e7curve}.

Again, due to the same commutation relations \eqref{calcr}, when we have the asymptotic value $q^{\frac{n}{2}}$ for large $\widehat Q$ or $\widehat P$, the asymptotic value for small $\widehat Q$ or $\widehat P$ is $q^{-\frac{n}{2}}$.
For this reason, the asymptotic values in table \ref{e7asymptotic} can always be paired as
\begin{align}
f_j\cdot(g_1h_j)^{-1}=1,\;(j=1,\cdots,4),\quad g_1\cdot g_2=1.
\label{antipode}
\end{align}

Note that, as discussed in \cite{KM}, since generally similarity transformations rotate the operators $\widehat{\cal Q}$ and $\widehat{\cal P}$ cyclically and change the asymptotic values, after we fix the reference interval where the Hanany-Witten transitions are forbidden, we allow only ``small'' similarity transformations \eqref{AB}.
Also, since each pair of asymptotic values on the opposite sides corresponds to each 5-brane, we distinguish 5-branes by labels.
Here we label $(1,k)$5-branes by Arabic numerals $1,2,3,4$ and NS5-branes by Roman numerals $\text{i},\text{ii}$.
Note that, while four Arabic numerals for the four $(1,k)$5-branes correspond to the label for the pairs $\{f_j,(g_1h_j)^{-1}\}_{j=1,2,3,4}$ of horizontal asymptotic values of $\widehat Q$, two Roman numerals for the two NS5-branes are not related respectively to $g_1$ and $g_2$.
Instead $\{g_1,g_2\}$ forms a pair of doubly degenerate vertical asymptotic values of $\widehat P$ and corresponds to the combination of the two NS5-branes separated by two $(1,k)$5-branes.
Due to this reason, for the time being, we always place the operators $\widehat{\cal P}_\text{i}$ on the left and $\widehat{\cal P}_\text{ii}$ on the right.
We will come back to the exchange of the two operators $\widehat{\cal P}$ later.
After further fixing the condition \eqref{e7gauge}, from the asymptotic values given in table \ref{e7asymptotic} we can read off the parameters explicitly for each ordering of $(1,k)$5-branes.
The result is given in table \ref{e7parameter}.

\begin{table}[t!]
\begin{center}
\begin{tabular}{l||l|l}
type&spectral operator&$(f_1,f_2,f_3,g_1,h_1,h_2,h_3)$\\\hline\hline
$\widehat{\cal Q}^2\widehat{\cal P}\widehat{\cal Q}^2\widehat{\cal P}$
&$\widehat{\cal Q}_4\widehat{\cal Q}_3\widehat{\cal P}_\text{i}\widehat{\cal Q}_2\widehat{\cal Q}_1\widehat{\cal P}_\text{ii}$
&$(q^{\frac{1}{2}},q^{\frac{1}{2}},1,q^{-1},q^{\frac{1}{2}},q^{\frac{1}{2}},1)$\\\cline{2-3}
&$\widehat{\cal Q}_4\widehat{\cal Q}_2\widehat{\cal P}_\text{i}\widehat{\cal Q}_3\widehat{\cal Q}_1\widehat{\cal P}_\text{ii}$
&$(q^{\frac{1}{2}},1,q^{\frac{1}{2}},q^{-1},q^{\frac{1}{2}},1,q^{\frac{1}{2}})$\\\cline{2-3}
&$\widehat{\cal Q}_4\widehat{\cal Q}_1\widehat{\cal P}_\text{i}\widehat{\cal Q}_3\widehat{\cal Q}_2\widehat{\cal P}_\text{ii}$
&$(1,q^{\frac{1}{2}},q^{\frac{1}{2}},q^{-1},1,q^{\frac{1}{2}},q^{\frac{1}{2}})$\\\cline{2-3}
&$\widehat{\cal Q}_3\widehat{\cal Q}_2\widehat{\cal P}_\text{i}\widehat{\cal Q}_4\widehat{\cal Q}_1\widehat{\cal P}_\text{ii}$
&$(1,q^{-\frac{1}{2}},q^{-\frac{1}{2}},q^{-1},q,q^{\frac{1}{2}},q^{\frac{1}{2}})$\\\cline{2-3}
&$\widehat{\cal Q}_3\widehat{\cal Q}_1\widehat{\cal P}_\text{i}\widehat{\cal Q}_4\widehat{\cal Q}_2\widehat{\cal P}_\text{ii}$
&$(q^{-\frac{1}{2}},1,q^{-\frac{1}{2}},q^{-1},q^{\frac{1}{2}},q,q^{\frac{1}{2}})$\\\cline{2-3}
&$\widehat{\cal Q}_2\widehat{\cal Q}_1\widehat{\cal P}_\text{i}\widehat{\cal Q}_4\widehat{\cal Q}_3\widehat{\cal P}_\text{ii}$
&$(q^{-\frac{1}{2}},q^{-\frac{1}{2}},1,q^{-1},q^{\frac{1}{2}},q^{\frac{1}{2}},q)$\\\hline\hline
$\widehat{\cal Q}\widehat{\cal P}\widehat{\cal Q}^2\widehat{\cal P}\widehat{\cal Q}$
&$\widehat{\cal Q}_4\widehat{\cal P}_\text{i}\widehat{\cal Q}_3\widehat{\cal Q}_2\widehat{\cal P}_\text{ii}\widehat{\cal Q}_1$
&$(q,q^{\frac{1}{2}},q^{\frac{1}{2}},1,1,q^{-\frac{1}{2}},q^{-\frac{1}{2}})$\\\cline{2-3}
&$\widehat{\cal Q}_4\widehat{\cal P}_\text{i}\widehat{\cal Q}_3\widehat{\cal Q}_1\widehat{\cal P}_\text{ii}\widehat{\cal Q}_2$
&$(q^{\frac{1}{2}},q,q^{\frac{1}{2}},1,q^{-\frac{1}{2}},1,q^{-\frac{1}{2}})$\\\cline{2-3}
&$\widehat{\cal Q}_4\widehat{\cal P}_\text{i}\widehat{\cal Q}_2\widehat{\cal Q}_1\widehat{\cal P}_\text{ii}\widehat{\cal Q}_3$
&$(q^{\frac{1}{2}},q^{\frac{1}{2}},q,1,q^{-\frac{1}{2}},q^{-\frac{1}{2}},1)$\\\cline{2-3}
&$\widehat{\cal Q}_3\widehat{\cal P}_\text{i}\widehat{\cal Q}_4\widehat{\cal Q}_2\widehat{\cal P}_\text{ii}\widehat{\cal Q}_1$
&$(q^{\frac{1}{2}},1,q^{-\frac{1}{2}},1,q^{\frac{1}{2}},1,q^{-\frac{1}{2}})$\\\cline{2-3}
&$\widehat{\cal Q}_3\widehat{\cal P}_\text{i}\widehat{\cal Q}_4\widehat{\cal Q}_1\widehat{\cal P}_\text{ii}\widehat{\cal Q}_2$
&$(1,q^{\frac{1}{2}},q^{-\frac{1}{2}},1,1,q^{\frac{1}{2}},q^{-\frac{1}{2}})$\\\cline{2-3}
&$\widehat{\cal Q}_3\widehat{\cal P}_\text{i}\widehat{\cal Q}_2\widehat{\cal Q}_1\widehat{\cal P}_\text{ii}\widehat{\cal Q}_4$
&$(q^{-\frac{1}{2}},q^{-\frac{1}{2}},q^{-1},1,q^{\frac{1}{2}},q^{\frac{1}{2}},1)$\\\cline{2-3}
&$\widehat{\cal Q}_2\widehat{\cal P}_\text{i}\widehat{\cal Q}_4\widehat{\cal Q}_3\widehat{\cal P}_\text{ii}\widehat{\cal Q}_1$
&$(q^{\frac{1}{2}},q^{-\frac{1}{2}},1,1,q^{\frac{1}{2}},q^{-\frac{1}{2}},1)$\\\cline{2-3}
&$\widehat{\cal Q}_2\widehat{\cal P}_\text{i}\widehat{\cal Q}_4\widehat{\cal Q}_1\widehat{\cal P}_\text{ii}\widehat{\cal Q}_3$
&$(1,q^{-\frac{1}{2}},q^{\frac{1}{2}},1,1,q^{-\frac{1}{2}},q^{\frac{1}{2}})$\\\cline{2-3}
&$\widehat{\cal Q}_2\widehat{\cal P}_\text{i}\widehat{\cal Q}_3\widehat{\cal Q}_1\widehat{\cal P}_\text{ii}\widehat{\cal Q}_4$
&$(q^{-\frac{1}{2}},q^{-1},q^{-\frac{1}{2}},1,q^{\frac{1}{2}},1,q^{\frac{1}{2}})$\\\cline{2-3}
&$\widehat{\cal Q}_1\widehat{\cal P}_\text{i}\widehat{\cal Q}_4\widehat{\cal Q}_3\widehat{\cal P}_\text{ii}\widehat{\cal Q}_2$
&$(q^{-\frac{1}{2}},q^{\frac{1}{2}},1,1,q^{-\frac{1}{2}},q^{\frac{1}{2}},1)$\\\cline{2-3}
&$\widehat{\cal Q}_1\widehat{\cal P}_\text{i}\widehat{\cal Q}_4\widehat{\cal Q}_2\widehat{\cal P}_\text{ii}\widehat{\cal Q}_3$
&$(q^{-\frac{1}{2}},1,q^{\frac{1}{2}},1,q^{-\frac{1}{2}},1,q^{\frac{1}{2}})$\\\cline{2-3}
&$\widehat{\cal Q}_1\widehat{\cal P}_\text{i}\widehat{\cal Q}_3\widehat{\cal Q}_2\widehat{\cal P}_\text{ii}\widehat{\cal Q}_4$
&$(q^{-1},q^{-\frac{1}{2}},q^{-\frac{1}{2}},1,1,q^{\frac{1}{2}},q^{\frac{1}{2}})$\\\hline\hline
$\widehat{\cal P}\widehat{\cal Q}^2\widehat{\cal P}\widehat{\cal Q}^2$
&$\widehat{\cal P}_\text{i}\widehat{\cal Q}_4\widehat{\cal Q}_3\widehat{\cal P}_\text{ii}\widehat{\cal Q}_2\widehat{\cal Q}_1$
&$(q^{\frac{1}{2}},q^{\frac{1}{2}},1,q,q^{-\frac{1}{2}},q^{-\frac{1}{2}},q^{-1})$\\\cline{2-3}
&$\widehat{\cal P}_\text{i}\widehat{\cal Q}_4\widehat{\cal Q}_2\widehat{\cal P}_\text{ii}\widehat{\cal Q}_3\widehat{\cal Q}_1$
&$(q^{\frac{1}{2}},1,q^{\frac{1}{2}},q,q^{-\frac{1}{2}},q^{-1},q^{-\frac{1}{2}})$\\\cline{2-3}
&$\widehat{\cal P}_\text{i}\widehat{\cal Q}_4\widehat{\cal Q}_1\widehat{\cal P}_\text{ii}\widehat{\cal Q}_3\widehat{\cal Q}_2$
&$(1,q^{\frac{1}{2}},q^{\frac{1}{2}},q,q^{-1},q^{-\frac{1}{2}},q^{-\frac{1}{2}})$\\\cline{2-3}
&$\widehat{\cal P}_\text{i}\widehat{\cal Q}_3\widehat{\cal Q}_2\widehat{\cal P}_\text{ii}\widehat{\cal Q}_4\widehat{\cal Q}_1$
&$(1,q^{-\frac{1}{2}},q^{-\frac{1}{2}},q,1,q^{-\frac{1}{2}},q^{-\frac{1}{2}})$\\\cline{2-3}
&$\widehat{\cal P}_\text{i}\widehat{\cal Q}_3\widehat{\cal Q}_1\widehat{\cal P}_\text{ii}\widehat{\cal Q}_4\widehat{\cal Q}_2$
&$(q^{-\frac{1}{2}},1,q^{-\frac{1}{2}},q,q^{-\frac{1}{2}},1,q^{-\frac{1}{2}})$\\\cline{2-3}
&$\widehat{\cal P}_\text{i}\widehat{\cal Q}_2\widehat{\cal Q}_1\widehat{\cal P}_\text{ii}\widehat{\cal Q}_4\widehat{\cal Q}_3$
&$(q^{-\frac{1}{2}},q^{-\frac{1}{2}},1,q,q^{-\frac{1}{2}},q^{-\frac{1}{2}},1)$
\end{tabular}
\end{center}
\caption{The parameters of the $E_7$ quantum curve for each ordering of branes.}
\label{e7parameter}
\end{table}

The first observation is again that all the parameters of the $E_7$ curve corresponding to brane configurations are located in the four-dimensional subspace
\begin{align}
{\cal C}_\text{B}^{2,4}\cap{\cal C}_\text{P}^{E_7}=\{(f_1,f_2,f_3,g_1)\}
=\biggl\{\biggl(f_1,f_2,f_3,g_1,\sqrt{\frac{f_1}{f_2f_3g_1}},\sqrt{\frac{f_2}{f_1f_3g_1}},\sqrt{\frac{f_3}{f_1f_2g_1}}\biggr)\biggr\}.
\label{e7sub}
\end{align}
Although it was simply an observation in \cite{KM}, here we can provide an intuitive explanation for the subspace.
Due to the commutation relations \eqref{calcr} the asymptotic values on the opposite sides are correlated as in \eqref{antipode}.
If we apply the relations \eqref{antipode} for the asymptotic values before fixing the condition \eqref{e7gauge}, we obtain
\begin{align}
&Af_1\cdot A(g_1h_1)^{-1}=1,\quad
Af_2\cdot A(g_1h_2)^{-1}=1,\quad
Af_3\cdot A(g_1h_3)^{-1}=1,\nonumber\\
&A\cdot A(f_1f_2f_3g_1h_1h_2h_3)=1,\quad
B\cdot Bg_1=1.
\end{align}
After eliminating $A$, it is clear that $h_j$ ($j=1,2,3$) is expressed in terms of $(f_1,f_2,f_3,g_1)$ as in \eqref{e7sub}.
For the $D_5$ case in the previous section, the fact \eqref{hheee} that the space of brane configurations ${\cal C}_\text{B}^{2,2}$ is located in the three-dimensional subspace of the five-dimensional space of point configurations ${\cal C}_\text{P}^{D_5}$ can be explained by the same arguments.

For keen readers, the counting of dimensions may already be clear in \eqref{antipode} even before fixing the condition \eqref{e7gauge}.
Namely before the condition \eqref{e7gauge} we have ten parameters $(f_1,f_2,f_3,f_4,g_1,g_2,h_1,h_2,h_3,h_4)$ subject to one constraint \eqref{e7constraint} leaving nine degrees of freedom.
For the interpretation in brane configurations, the asymptotic values on the opposite sides should be paired as in \eqref{antipode}, which gives five relations.
By subtraction, only four dimensions remain, which matches with $\dim{\cal C}_\text{B}^{2,4}\cap{\cal C}_\text{P}^{E_7}=4$.
Similarly, for the $D_5$ case in the previous section, eight asymptotic values in figure \ref{d5asympt} subject to one constraint leave seven degrees of freedom.
After subtracting four relations coming from pairing the asymptotic values on the opposite sides for the brane interpretation, we find three dimensions matching with $\dim{\cal C}_\text{B}^{2,2}=3$.

To identify the point configurations for the $E_7$ curve with the relative ranks of the brane configurations, let us list all the cases.
For the point configurations we regard the case of $\widehat{\cal Q}_4\widehat{\cal Q}_3\widehat{\cal P}_\text{i}\widehat{\cal Q}_2\widehat{\cal Q}_1\widehat{\cal P}_\text{ii}$ as the reference and consider the parameters relative to it
\begin{align}
\delta(f_1,f_2,f_3,g_1)=\frac{(f_1,f_2,f_3,g_1)}
{(f_1,f_2,f_3,g_1)|_{\stackrel{\text{ii}}{\bullet}\,\stackrel{1}{\circ}\,\stackrel{2}{\circ}\,\stackrel{\text{i}}{\bullet}\,\stackrel{3}{\circ}\,\stackrel{4}{\circ}}},
\label{deltafg}
\end{align}
(where the division applies to each component separately) with
\begin{align}
(f_1,f_2,f_3,g_1)|_{\stackrel{\text{ii}}{\bullet}\,\stackrel{1}{\circ}\,\stackrel{2}{\circ}\,\stackrel{\text{i}}{\bullet}\,\stackrel{3}{\circ}\,\stackrel{4}{\circ}}
=(q^{\frac{1}{2}},q^{\frac{1}{2}},1,q^{-1}).
\end{align}
For the brane configurations, we rewrite those without rank deformations into the standard order $\langle\stackrel
{\text{ii}}{\bullet}\,\stackrel{1}{\circ}\,\stackrel{2}{\circ}\,\stackrel{\text{i}}{\bullet}\,\stackrel{3}{\circ}\,\stackrel{4}{\circ}\rangle$ by applying the Hanany-Witten transitions \eqref{HW}.
We summarize the result in table \ref{comparison}.

\begin{table}[t!]
\begin{center}
\begin{tabular}{l|l||l}
spectral operator&$\delta(f_1,f_2,f_3,g_1)$&brane configuration\\\hline\hline
$\widehat{\cal Q}_4\widehat{\cal Q}_3\widehat{\cal P}_\text{i}\widehat{\cal Q}_2\widehat{\cal Q}_1\widehat{\cal P}_\text{ii}$
&$(1,1,1,1)$
&$\langle\stackrel{\text{ii}}{\bullet}\,\stackrel{1}{\circ}\,\stackrel{2}{\circ}\,\stackrel{\text{i}}{\bullet}\,\stackrel{3}{\circ}\,\stackrel{4}{\circ}\rangle
=\langle 0\stackrel{\text{ii}}{\bullet}0\stackrel{1}{\circ}0\stackrel{2}{\circ}0\stackrel{\text{i}}{\bullet}0\stackrel{3}{\circ}0\stackrel{4}{\circ}\rangle$
\\\hline
$\widehat{\cal Q}_4\widehat{\cal Q}_2\widehat{\cal P}_\text{i}\widehat{\cal Q}_3\widehat{\cal Q}_1\widehat{\cal P}_\text{ii}$
&$(1,q^{-\frac{1}{2}},q^{\frac{1}{2}},1)$
&$\langle\stackrel{\text{ii}}{\bullet}\,\stackrel{1}{\circ}\,\stackrel{3}{\circ}\,\stackrel{\text{i}}{\bullet}\,\stackrel{2}{\circ}\,\stackrel{4}{\circ}\rangle
=\langle 0\stackrel{\text{ii}}{\bullet}0\stackrel{1}{\circ}0\stackrel{2}{\circ}k\stackrel{\text{i}}{\bullet}k\stackrel{3}{\circ}0\stackrel{4}{\circ}\rangle$
\\\hline
$\widehat{\cal Q}_4\widehat{\cal Q}_1\widehat{\cal P}_\text{i}\widehat{\cal Q}_3\widehat{\cal Q}_2\widehat{\cal P}_\text{ii}$
&$(q^{-\frac{1}{2}},1,q^{\frac{1}{2}},1)$
&$\langle\stackrel{\text{ii}}{\bullet}\,\stackrel{2}{\circ}\,\stackrel{3}{\circ}\,\stackrel{\text{i}}{\bullet}\,\stackrel{1}{\circ}\,\stackrel{4}{\circ}\rangle
=\langle 0\stackrel{\text{ii}}{\bullet}0\stackrel{1}{\circ}k\stackrel{2}{\circ}k\stackrel{\text{i}}{\bullet}k\stackrel{3}{\circ}0\stackrel{4}{\circ}\rangle$
\\\hline
$\widehat{\cal Q}_3\widehat{\cal Q}_2\widehat{\cal P}_\text{i}\widehat{\cal Q}_4\widehat{\cal Q}_1\widehat{\cal P}_\text{ii}$
&$(q^{-\frac{1}{2}},q^{-1},q^{-\frac{1}{2}},1)$
&$\langle\stackrel{\text{ii}}{\bullet}\,\stackrel{1}{\circ}\,\stackrel{4}{\circ}\,\stackrel{\text{i}}{\bullet}\,\stackrel{2}{\circ}\,\stackrel{3}{\circ}\rangle
=\langle 0\stackrel{\text{ii}}{\bullet}0\stackrel{1}{\circ}0\stackrel{2}{\circ}k\stackrel{\text{i}}{\bullet}k\stackrel{3}{\circ}k\stackrel{4}{\circ}\rangle$
\\\hline
$\widehat{\cal Q}_3\widehat{\cal Q}_1\widehat{\cal P}_\text{i}\widehat{\cal Q}_4\widehat{\cal Q}_2\widehat{\cal P}_\text{ii}$
&$(q^{-1},q^{-\frac{1}{2}},q^{-\frac{1}{2}},1)$
&$\langle\stackrel{\text{ii}}{\bullet}\,\stackrel{2}{\circ}\,\stackrel{4}{\circ}\,\stackrel{\text{i}}{\bullet}\,\stackrel{1}{\circ}\,\stackrel{3}{\circ}\rangle
=\langle 0\stackrel{\text{ii}}{\bullet}0\stackrel{1}{\circ}k\stackrel{2}{\circ}k\stackrel{\text{i}}{\bullet}k\stackrel{3}{\circ}k\stackrel{4}{\circ}\rangle$
\\\hline
$\widehat{\cal Q}_2\widehat{\cal Q}_1\widehat{\cal P}_\text{i}\widehat{\cal Q}_4\widehat{\cal Q}_3\widehat{\cal P}_\text{ii}$
&$(q^{-1},q^{-1},1,1)$
&$\langle\stackrel{\text{ii}}{\bullet}\,\stackrel{3}{\circ}\,\stackrel{4}{\circ}\,\stackrel{\text{i}}{\bullet}\,\stackrel{1}{\circ}\,\stackrel{2}{\circ}\rangle
=\langle 0\stackrel{\text{ii}}{\bullet}0\stackrel{1}{\circ}k\stackrel{2}{\circ}2k\stackrel{\text{i}}{\bullet}2k\stackrel{3}{\circ}k\stackrel{4}{\circ}\rangle$
\\\hline\hline
$\widehat{\cal Q}_4\widehat{\cal P}_\text{i}\widehat{\cal Q}_3\widehat{\cal Q}_2\widehat{\cal P}_\text{ii}\widehat{\cal Q}_1$
&$(q^{\frac{1}{2}},1,q^{\frac{1}{2}},q)$
&$\langle\stackrel{1}{\circ}\,\stackrel{\text{ii}}{\bullet}\,\stackrel{2}{\circ}\,\stackrel{3}{\circ}\,\stackrel{\text{i}}{\bullet}\,\stackrel{4}{\circ}\rangle
=\langle 0\stackrel{\text{ii}}{\bullet}k\stackrel{1}{\circ}0\stackrel{2}{\circ}0\stackrel{\text{i}}{\bullet}k\stackrel{3}{\circ}0\stackrel{4}{\circ}\rangle$
\\\hline
$\widehat{\cal Q}_4\widehat{\cal P}_\text{i}\widehat{\cal Q}_3\widehat{\cal Q}_1\widehat{\cal P}_\text{ii}\widehat{\cal Q}_2$
&$(1,q^{\frac{1}{2}},q^{\frac{1}{2}},q)$
&$\langle\stackrel{2}{\circ}\,\stackrel{\text{ii}}{\bullet}\,\stackrel{1}{\circ}\,\stackrel{3}{\circ}\,\stackrel{\text{i}}{\bullet}\,\stackrel{4}{\circ}\rangle
=\langle 0\stackrel{\text{ii}}{\bullet}k\stackrel{1}{\circ}k\stackrel{2}{\circ}0\stackrel{\text{i}}{\bullet}k\stackrel{3}{\circ}0\stackrel{4}{\circ}\rangle$
\\\hline
$\widehat{\cal Q}_4\widehat{\cal P}_\text{i}\widehat{\cal Q}_2\widehat{\cal Q}_1\widehat{\cal P}_\text{ii}\widehat{\cal Q}_3$
&$(1,1,q,q)$
&$\langle\stackrel{3}{\circ}\,\stackrel{\text{ii}}{\bullet}\,\stackrel{1}{\circ}\,\stackrel{2}{\circ}\,\stackrel{\text{i}}{\bullet}\,\stackrel{4}{\circ}\rangle
=\langle 0\stackrel{\text{ii}}{\bullet}k\stackrel{1}{\circ}k\stackrel{2}{\circ}k\stackrel{\text{i}}{\bullet}2k\stackrel{3}{\circ}0\stackrel{4}{\circ}\rangle$
\\\hline
$\widehat{\cal Q}_3\widehat{\cal P}_\text{i}\widehat{\cal Q}_4\widehat{\cal Q}_2\widehat{\cal P}_\text{ii}\widehat{\cal Q}_1$
&$(1,q^{-\frac{1}{2}},q^{-\frac{1}{2}},q)$
&$\langle\stackrel{1}{\circ}\,\stackrel{\text{ii}}{\bullet}\,\stackrel{2}{\circ}\,\stackrel{4}{\circ}\,\stackrel{\text{i}}{\bullet}\,\stackrel{3}{\circ}\rangle
=\langle 0\stackrel{\text{ii}}{\bullet}k\stackrel{1}{\circ}0\stackrel{2}{\circ}0\stackrel{\text{i}}{\bullet}k\stackrel{3}{\circ}k\stackrel{4}{\circ}\rangle$
\\\hline
$\widehat{\cal Q}_3\widehat{\cal P}_\text{i}\widehat{\cal Q}_4\widehat{\cal Q}_1\widehat{\cal P}_\text{ii}\widehat{\cal Q}_2$
&$(q^{-\frac{1}{2}},1,q^{-\frac{1}{2}},q)$
&$\langle\stackrel{2}{\circ}\,\stackrel{\text{ii}}{\bullet}\,\stackrel{1}{\circ}\,\stackrel{4}{\circ}\,\stackrel{\text{i}}{\bullet}\,\stackrel{3}{\circ}\rangle
=\langle 0\stackrel{\text{ii}}{\bullet}k\stackrel{1}{\circ}k\stackrel{2}{\circ}0\stackrel{\text{i}}{\bullet}k\stackrel{3}{\circ}k\stackrel{4}{\circ}\rangle$
\\\hline
$\widehat{\cal Q}_3\widehat{\cal P}_\text{i}\widehat{\cal Q}_2\widehat{\cal Q}_1\widehat{\cal P}_\text{ii}\widehat{\cal Q}_4$
&$(q^{-1},q^{-1},q^{-1},q)$
&$\langle\stackrel{4}{\circ}\,\stackrel{\text{ii}}{\bullet}\,\stackrel{1}{\circ}\,\stackrel{2}{\circ}\,\stackrel{\text{i}}{\bullet}\,\stackrel{3}{\circ}\rangle
=\langle 0\stackrel{\text{ii}}{\bullet}k\stackrel{1}{\circ}k\stackrel{2}{\circ}k\stackrel{\text{i}}{\bullet}2k\stackrel{3}{\circ}2k\stackrel{4}{\circ}\rangle$
\\\hline
$\widehat{\cal Q}_2\widehat{\cal P}_\text{i}\widehat{\cal Q}_4\widehat{\cal Q}_3\widehat{\cal P}_\text{ii}\widehat{\cal Q}_1$
&$(1,q^{-1},1,q)$
&$\langle\stackrel{1}{\circ}\,\stackrel{\text{ii}}{\bullet}\,\stackrel{3}{\circ}\,\stackrel{4}{\circ}\,\stackrel{\text{i}}{\bullet}\,\stackrel{2}{\circ}\rangle
=\langle 0\stackrel{\text{ii}}{\bullet}k\stackrel{1}{\circ}0\stackrel{2}{\circ}k\stackrel{\text{i}}{\bullet}2k\stackrel{3}{\circ}k\stackrel{4}{\circ}\rangle$
\\\hline
$\widehat{\cal Q}_2\widehat{\cal P}_\text{i}\widehat{\cal Q}_4\widehat{\cal Q}_1\widehat{\cal P}_\text{ii}\widehat{\cal Q}_3$
&$(q^{-\frac{1}{2}},q^{-1},q^{\frac{1}{2}},q)$
&$\langle\stackrel{3}{\circ}\,\stackrel{\text{ii}}{\bullet}\,\stackrel{1}{\circ}\,\stackrel{4}{\circ}\,\stackrel{\text{i}}{\bullet}\,\stackrel{2}{\circ}\rangle
=\langle 0\stackrel{\text{ii}}{\bullet}k\stackrel{1}{\circ}k\stackrel{2}{\circ}2k\stackrel{\text{i}}{\bullet}3k\stackrel{3}{\circ}k\stackrel{4}{\circ}\rangle$
\\\hline
$\widehat{\cal Q}_2\widehat{\cal P}_\text{i}\widehat{\cal Q}_3\widehat{\cal Q}_1\widehat{\cal P}_\text{ii}\widehat{\cal Q}_4$
&$(q^{-1},q^{-\frac{3}{2}},q^{-\frac{1}{2}},q)$
&$\langle\stackrel{4}{\circ}\,\stackrel{\text{ii}}{\bullet}\,\stackrel{1}{\circ}\,\stackrel{3}{\circ}\,\stackrel{\text{i}}{\bullet}\,\stackrel{2}{\circ}\rangle
=\langle 0\stackrel{\text{ii}}{\bullet}k\stackrel{1}{\circ}k\stackrel{2}{\circ}2k\stackrel{\text{i}}{\bullet}3k\stackrel{3}{\circ}2k\stackrel{4}{\circ}\rangle$
\\\hline
$\widehat{\cal Q}_1\widehat{\cal P}_\text{i}\widehat{\cal Q}_4\widehat{\cal Q}_3\widehat{\cal P}_\text{ii}\widehat{\cal Q}_2$
&$(q^{-1},1,1,q)$
&$\langle\stackrel{2}{\circ}\,\stackrel{\text{ii}}{\bullet}\,\stackrel{3}{\circ}\,\stackrel{4}{\circ}\,\stackrel{\text{i}}{\bullet}\,\stackrel{1}{\circ}\rangle
=\langle 0\stackrel{\text{ii}}{\bullet}k\stackrel{1}{\circ}2k\stackrel{2}{\circ}k\stackrel{\text{i}}{\bullet}2k\stackrel{3}{\circ}k\stackrel{4}{\circ}\rangle$
\\\hline
$\widehat{\cal Q}_1\widehat{\cal P}_\text{i}\widehat{\cal Q}_4\widehat{\cal Q}_2\widehat{\cal P}_\text{ii}\widehat{\cal Q}_3$
&$(q^{-1},q^{-\frac{1}{2}},q^{\frac{1}{2}},q)$
&$\langle\stackrel{3}{\circ}\,\stackrel{\text{ii}}{\bullet}\,\stackrel{2}{\circ}\,\stackrel{4}{\circ}\,\stackrel{\text{i}}{\bullet}\,\stackrel{1}{\circ}\rangle
=\langle 0\stackrel{\text{ii}}{\bullet}k\stackrel{1}{\circ}2k\stackrel{2}{\circ}2k\stackrel{\text{i}}{\bullet}3k\stackrel{3}{\circ}k\stackrel{4}{\circ}\rangle$
\\\hline
$\widehat{\cal Q}_1\widehat{\cal P}_\text{i}\widehat{\cal Q}_3\widehat{\cal Q}_2\widehat{\cal P}_\text{ii}\widehat{\cal Q}_4$
&$(q^{-\frac{3}{2}},q^{-1},q^{-\frac{1}{2}},q)$
&$\langle\stackrel{4}{\circ}\,\stackrel{\text{ii}}{\bullet}\,\stackrel{2}{\circ}\,\stackrel{3}{\circ}\,\stackrel{\text{i}}{\bullet}\,\stackrel{1}{\circ}\rangle
=\langle 0\stackrel{\text{ii}}{\bullet}k\stackrel{1}{\circ}2k\stackrel{2}{\circ}2k\stackrel{\text{i}}{\bullet}3k\stackrel{3}{\circ}2k\stackrel{4}{\circ}\rangle$
\\\hline\hline
$\widehat{\cal P}_\text{i}\widehat{\cal Q}_4\widehat{\cal Q}_3\widehat{\cal P}_\text{ii}\widehat{\cal Q}_2\widehat{\cal Q}_1$
&$(1,1,1,q^2)$
&$\langle\stackrel{1}{\circ}\,\stackrel{2}{\circ}\,\stackrel{\text{ii}}{\bullet}\,\stackrel{3}{\circ}\,\stackrel{4}{\circ}\,\stackrel{\text{i}}{\bullet}\rangle
=\langle 0\stackrel{\text{ii}}{\bullet}2k\stackrel{1}{\circ}k\stackrel{2}{\circ}0\stackrel{\text{i}}{\bullet}2k\stackrel{3}{\circ}k\stackrel{4}{\circ}\rangle$
\\\hline
$\widehat{\cal P}_\text{i}\widehat{\cal Q}_4\widehat{\cal Q}_2\widehat{\cal P}_\text{ii}\widehat{\cal Q}_3\widehat{\cal Q}_1$
&$(1,q^{-\frac{1}{2}},q^{\frac{1}{2}},q^2)$
&$\langle\stackrel{1}{\circ}\,\stackrel{3}{\circ}\,\stackrel{\text{ii}}{\bullet}\,\stackrel{2}{\circ}\,\stackrel{4}{\circ}\,\stackrel{\text{i}}{\bullet}\rangle
=\langle 0\stackrel{\text{ii}}{\bullet}2k\stackrel{1}{\circ}k\stackrel{2}{\circ}k\stackrel{\text{i}}{\bullet}3k\stackrel{3}{\circ}k\stackrel{4}{\circ}\rangle$
\\\hline
$\widehat{\cal P}_\text{i}\widehat{\cal Q}_4\widehat{\cal Q}_1\widehat{\cal P}_\text{ii}\widehat{\cal Q}_3\widehat{\cal Q}_2$
&$(q^{-\frac{1}{2}},1,q^{\frac{1}{2}},q^2)$
&$\langle\stackrel{2}{\circ}\,\stackrel{3}{\circ}\,\stackrel{\text{ii}}{\bullet}\,\stackrel{1}{\circ}\,\stackrel{4}{\circ}\,\stackrel{\text{i}}{\bullet}\rangle
=\langle 0\stackrel{\text{ii}}{\bullet}2k\stackrel{1}{\circ}2k\stackrel{2}{\circ}k\stackrel{\text{i}}{\bullet}3k\stackrel{3}{\circ}k\stackrel{4}{\circ}\rangle$
\\\hline
$\widehat{\cal P}_\text{i}\widehat{\cal Q}_3\widehat{\cal Q}_2\widehat{\cal P}_\text{ii}\widehat{\cal Q}_4\widehat{\cal Q}_1$
&$(q^{-\frac{1}{2}},q^{-1},q^{-\frac{1}{2}},q^2)$
&$\langle\stackrel{1}{\circ}\,\stackrel{4}{\circ}\,\stackrel{\text{ii}}{\bullet}\,\stackrel{2}{\circ}\,\stackrel{3}{\circ}\,\stackrel{\text{i}}{\bullet}\rangle
=\langle 0\stackrel{\text{ii}}{\bullet}2k\stackrel{1}{\circ}k\stackrel{2}{\circ}k\stackrel{\text{i}}{\bullet}3k\stackrel{3}{\circ}2k\stackrel{4}{\circ}\rangle$
\\\hline
$\widehat{\cal P}_\text{i}\widehat{\cal Q}_3\widehat{\cal Q}_1\widehat{\cal P}_\text{ii}\widehat{\cal Q}_4\widehat{\cal Q}_2$
&$(q^{-1},q^{-\frac{1}{2}},q^{-\frac{1}{2}},q^2)$
&$\langle\stackrel{2}{\circ}\,\stackrel{4}{\circ}\,\stackrel{\text{ii}}{\bullet}\,\stackrel{1}{\circ}\,\stackrel{3}{\circ}\,\stackrel{\text{i}}{\bullet}\rangle
=\langle 0\stackrel{\text{ii}}{\bullet}2k\stackrel{1}{\circ}2k\stackrel{2}{\circ}k\stackrel{\text{i}}{\bullet}3k\stackrel{3}{\circ}2k\stackrel{4}{\circ}\rangle$
\\\hline
$\widehat{\cal P}_\text{i}\widehat{\cal Q}_2\widehat{\cal Q}_1\widehat{\cal P}_\text{ii}\widehat{\cal Q}_4\widehat{\cal Q}_3$
&$(q^{-1},q^{-1},1,q^2)$
&$\langle\stackrel{3}{\circ}\,\stackrel{4}{\circ}\,\stackrel{\text{ii}}{\bullet}\,\stackrel{1}{\circ}\,\stackrel{2}{\circ}\,\stackrel{\text{i}}{\bullet}\rangle
=\langle 0\stackrel{\text{ii}}{\bullet}2k\stackrel{1}{\circ}2k\stackrel{2}{\circ}2k\stackrel{\text{i}}{\bullet}4k\stackrel{3}{\circ}2k\stackrel{4}{\circ}\rangle$
\end{tabular}
\end{center}
\caption{Comparison between the point configurations for the $E_7$ quantum curve and the brane configurations.}
\label{comparison}
\end{table}

From the comparison in table \ref{comparison}, it is not difficult to read off the relation between the relative parameters of the point configurations $(f_1,f_2,f_3,g_1)=(e^{2\pi iF_1},e^{2\pi iF_2},e^{2\pi iF_3},e^{2\pi iG_1})$ and the rank differences obtained after applying the Hanany-Witten transitions,
\begin{align}
&(e^{2\pi i\delta F_1},e^{2\pi i\delta F_2},e^{2\pi i\delta F_3},e^{2\pi i\delta G_1})
\Leftrightarrow
\Bigl\langle 0\stackrel{\text{ii}}{\bullet}\delta G_1\stackrel{1}{\circ}
\frac{1}{2}(-3\delta F_1+\delta F_2+\delta F_3+\delta G_1)\stackrel{2}{\circ}\nonumber\\
&-\delta F_1-\delta F_2+\delta F_3\stackrel{\text{i}}{\bullet}
-\delta F_1-\delta F_2+\delta F_3+\delta G_1\stackrel{3}{\circ}
\frac{1}{2}(-\delta F_1-\delta F_2-\delta F_3+\delta G_1)\stackrel{4}{\circ}\Bigr\rangle.
\end{align}
In other words, in the original variables instead of the relative ones, we find
\begin{align}
&(e^{2\pi iF_1},e^{2\pi iF_2},e^{2\pi iF_3},e^{2\pi iG_1})
\Leftrightarrow
\Bigl\langle 0\stackrel{\text{ii}}{\bullet}
G_1+k\stackrel{1}{\circ}
\frac{1}{2}(-3F_1+F_2+F_3+G_1)+k\stackrel{2}{\circ}\nonumber\\
&-F_1-F_2+F_3+k\stackrel{\text{i}}{\bullet}
-F_1-F_2+F_3+G_1+2k\stackrel{3}{\circ}
\frac{1}{2}(-F_1-F_2-F_3+G_1)+k\stackrel{4}{\circ}\Bigr\rangle,
\label{F123G1}
\end{align}
where we have omitted the overall rank $N$ which is summed up in the grand canonical ensemble.
Although the relation \eqref{F123G1} is read off directly from the cases without rank deformations in table \ref{comparison}, it is natural to extrapolate the relation \eqref{F123G1} to the whole space ${\cal C}_\text{B}^{2,4}\cap{\cal C}_\text{P}^{E_7}$ linearly and expect \eqref{F123G1} to be valid there as well.
Again, unlike the case in the previous section with ${\cal C}_\text{B}^{2,2}\subset{\cal C}_\text{P}^{D_5}$, note here that ${\cal C}_\text{B}^{2,4}$ is not located inside ${\cal C}_\text{P}^{E_7}$, namely, ${\cal C}_\text{B}^{2,4}\not\subset{\cal C}_\text{P}^{E_7}$.

If we identify the four-dimensional subspace of brane configurations ${\cal C}_\text{B}^{2,4}\cap{\cal C}_\text{P}^{E_7}$ in the five-dimensional space of relative ranks ${\cal C}_\text{B}^{2,4}$ by a constraint, the ranks should satisfy the relation
\begin{align}
{\cal C}_\text{B}^{2,4}\cap{\cal C}_\text{P}^{E_7}
=\{\langle N_1\stackrel{\text{ii}}{\bullet}N_2\stackrel{1}{\circ}N_3\stackrel{2}{\circ}
N_4\stackrel{\text{i}}{\bullet}N_5\stackrel{3}{\circ}N_6\stackrel{4}{\circ}\rangle|
N_1+N_5=N_2+N_4\},
\label{constraint}
\end{align}
which is unnecessary for the case in the previous section.
Inversely, in this subspace we can identify the rank differences by
\begin{align}
&F_1=(N_1+N_2-N_3-N_6+k)/2,\nonumber\\
&F_2=(N_1+N_3-N_4-N_6+k)/2=(N_2+N_3-N_5-N_6+k)/2,\nonumber\\
&F_3=(N_2+N_4-2N_6)/2=(N_1+N_5-2N_6)/2,\nonumber\\
&G_1=-N_1+N_2-k=-N_4+N_5-k,
\label{FGN}
\end{align}
where the expressions are not unique due to the relation \eqref{constraint}.
Interestingly, the relation $N_1+N_5=N_2+N_4$ in \eqref{constraint} can be understood as a balanced condition between both sides of the NS5-branes $\bullet$.
Note that the balanced condition depends on the number of the $(1,k)$5-branes $\circ$ between the two NS5-branes $\bullet$, such as
\begin{align}
&\{\langle N_1\stackrel{\text{ii}}{\bullet}N_2\stackrel{1}{\circ}N_3\stackrel{\text{i}}{\bullet}
N'_4\stackrel{2}{\circ}N_5\stackrel{3}{\circ}N_6\stackrel{4}{\circ}\rangle|
N_1+N'_4=N_2+N_3+k\},\nonumber\\
&\{\langle N_1\stackrel{\text{ii}}{\bullet}N_2\stackrel{1}{\circ}N_3\stackrel{2}{\circ}
N_4\stackrel{3}{\circ}N'_5\stackrel{\text{i}}{\bullet}N_6\stackrel{4}{\circ}\rangle|
N_1+N_6=N_2+N'_5-k\}.
\label{different1k5}
\end{align}

\subsection{Symmetries as ${\mathbb Z}_2$ folding}

So far we have identified the four-dimensional subspace of rank deformations for brane configurations ${\cal C}_\text{B}^{2,4}$ in the seven-dimensional space of point configurations ${\cal C}_\text{P}^{E_7}$.
Let us study the group structure of this subspace.

First we can ask which subgroup of the $E_7$ Weyl group acting on the four-dimensional subspace ${\cal C}_\text{B}^{2,4}\cap{\cal C}_\text{P}^{E_7}$ trivially.
In other words, we would like to know which elements in the $E_7$ Weyl group leave each point in the four-dimensional subspace invariant.
It turns out that there are eight elements forming $W((A_1)^3)$ which leaves each point invariant.
Here the generator for each $W(A_1)$ is
\begin{align}
&s_2s_1s_3s_2s_4s_3s_5s_6s_7s_6s_5s_3s_4s_2s_3s_1s_2:(f_1,f_2,f_3,g_1,h_1,h_2,h_3)\nonumber\\
&\qquad\qquad\qquad\mapsto(f_1,f_1f_2g_1h_2h_3,f_1f_3g_1h_2h_3,g_1,h_1,(f_1g_1h_3)^{-1},(f_1g_1h_2)^{-1}),\nonumber\\
&s_5s_1s_3s_2s_4s_3s_5s_6s_7s_6s_5s_3s_4s_2s_3s_1s_5:(f_1,f_2,f_3,g_1,h_1,h_2,h_3)\nonumber\\
&\qquad\qquad\qquad\mapsto(f_1f_2g_1h_1h_3,f_2,f_2f_3g_1h_1h_3,g_1,(f_2g_1h_3)^{-1},h_2,(f_2g_1h_1)^{-1}),\nonumber\\
&s_6s_5s_3s_2s_4s_3s_5s_6s_7s_6s_5s_3s_4s_2s_3s_5s_6:(f_1,f_2,f_3,g_1,h_1,h_2,h_3)\nonumber\\
&\qquad\qquad\qquad\mapsto(f_1f_3g_1h_1h_2,f_2f_3g_1h_1h_2,f_3,g_1,(f_3g_1h_2)^{-1},(f_3g_1h_1)^{-1},h_3).
\label{A13}
\end{align}

\begin{figure}[!t]
\centering\includegraphics[scale=0.5,angle=-90]{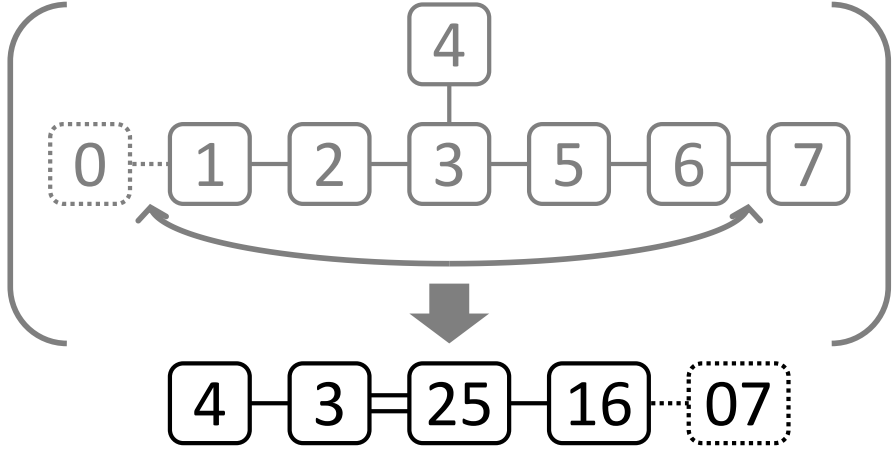}
\caption{Dynkin diagram of the $F_4$ algebra (denoting the symmetries for the four-dimensional subspace of brane configurations which can be realized by the $E_7$ quantum curve ${\cal C}_\text{B}^{2,4}\cap{\cal C}_\text{P}^{E_7}$).
This can be regarded as the ${\mathbb Z}_2$ folding of the $E_7$ Dynkin diagram.}
\label{f4alg}
\end{figure}

Furthermore, it is important to ask which subgroup of the $E_7$ Weyl group preserves the four-dimensional subspace ${\cal C}_\text{B}^{2,4}\cap{\cal C}_\text{P}^{E_7}$.
It turns out that, after modding out the trivial symmetries $W((A_1)^3)$ \eqref{A13}, $1152$ generators in the $E_7$ Weyl group containing $s_3$, $s_4$, $s_2s_5$, $s_1s_6$ and $s_0s_7$ leave the four-dimensional subspace invariant.
This is nothing but the $F_4$ Weyl group $W(F_4)$.
The $F_4$ Dynkin diagram is depicted in figure \ref{f4alg} and, again, this can be regarded as the ${\mathbb Z}_2$ folding of the $E_7$ Dynkin diagram.
This is understood as follows.
Since the commutation relation \eqref{calcr} requires the correlation of the opposite asymptotic values \eqref{antipode}, the transformations exchanging asymptotic values should be combined between the two acting on the opposite sides, $(f_1,f_2,f_3,1)$ and $((g_1h_1)^{-1},(g_1h_2)^{-1},(g_1h_3)^{-1},f_1f_2f_3g_1h_1h_2h_3)$.
For this reason, only the combined transformations
\begin{align}
s_2s_5&:(f_1,f_2,f_3,g_1,h_1,h_2,h_3)\mapsto(f_2,f_1,f_3,g_1,h_2,h_1,h_3),\nonumber\\
s_1s_6&:(f_1,f_2,f_3,g_1,h_1,h_2,h_3)\mapsto(f_1,f_3,f_2,g_1,h_1,h_3,h_2),\nonumber\\
s_0s_7&:(f_1,f_2,f_3,g_1,h_1,h_2,h_3)\mapsto(f_3^{-1}f_1,f_3^{-1}f_2,f_3^{-1},g_1,f_3h_1,f_3h_2,(f_1f_2g_1^2h_1h_2h_3)^{-1}),
\label{s251607}
\end{align}
remain as symmetries of the subspace ${\cal C}_\text{B}^{2,4}\cap{\cal C}_\text{P}^{E_7}$.
The fact that the invariant subgroup $W(B_3)$ is the ${\mathbb Z}_2$ folding of the original group $W(D_5)$ for the previous $D_5$ case can be explained by the same arguments.

As in the $D_5$ case, the ranks of $W((A_1)^3)$ and $W(F_4)$ are consistent with the dimensions of the perpendicular and parallel directions for the four-dimensional subspace of brane configurations ${\cal C}_\text{B}^{2,4}\cap{\cal C}_\text{P}^{E_7}$ in the seven-dimensional space of point configurations ${\cal C}_\text{P}^{E_7}$.

\subsection{Brane transitions}

After studying the symmetries for the four-dimensional space ${\cal C}_\text{B}^{2,4}\cap{\cal C}_\text{P}^{E_7}$ in the previous subsection, let us present an interpretation as brane transitions.

First the transformation $s_2s_5$ (and $s_1s_6$, $s_0s_7$ respectively) is interpreted as the Hanany-Witten transition exchanging the two $(1,k)$5-branes $\stackrel{1}{\circ}$/$\stackrel{2}{\circ}$ (and $\stackrel{2}{\circ}$/$\stackrel{3}{\circ}$, $\stackrel{3}{\circ}$/$\stackrel{4}{\circ}$ respectively).
For example, in exchanging the two $(1,k)$5-branes $\stackrel{1}{\circ}$/$\stackrel{2}{\circ}$, we obtain
\begin{align}
\langle N_1\stackrel{\text{ii}}{\bullet}N_2\stackrel{2}{\circ}N_3\stackrel{1}{\circ}
N_4\stackrel{\text{i}}{\bullet}N_5\stackrel{3}{\circ}N_6\stackrel{4}{\circ}\rangle
=\langle N_1\stackrel{\text{ii}}{\bullet}N_2\stackrel{1}{\circ}N'_3\stackrel{2}{\circ}
N_4\stackrel{\text{i}}{\bullet}N_5\stackrel{3}{\circ}N_6\stackrel{4}{\circ}\rangle,
\end{align}
with
\begin{align}
N'_3=N_2-N_3+N_4=\frac{1}{2}(F_1-3F_2+F_3+G_1)+k,
\end{align}
if we substitute \eqref{F123G1}.
Comparing with the original relation including $N_3=(-3F_1+F_2+F_3+G_1)/2+k$, we recognize the exchange of $F_1$ and $F_2$, which reproduce the combined transformation $s_2s_5$ \eqref{s251607}.
We can repeat similar computations for $s_1s_6$ and $s_0s_7$ as well.

The Hanany-Witten transition exchanging the two NS5-branes, however, is not a symmetry of ${\cal C}_\text{B}^{2,4}\cap{\cal C}_\text{P}^{E_7}$.
Instead, by exchanging the two NS5-branes $\stackrel{\text{i}}{\bullet}$/$\stackrel{\text{ii}}{\bullet}$, we find that the brane configuration \eqref{constraint} is transformed into
\begin{align}
\langle N_1\stackrel{\text{i}}{\bullet}N_2\stackrel{1}{\circ}N_3\stackrel{2}{\circ}
N_4\stackrel{\text{ii}}{\bullet}N_5\stackrel{3}{\circ}N_6\stackrel{4}{\circ}\rangle
=\langle N_1\stackrel{\text{ii}}{\bullet}N_2+2k\stackrel{1}{\circ}N_3+2k\stackrel{2}{\circ}
N_4+2k\stackrel{\text{i}}{\bullet}N_5\stackrel{3}{\circ}N_6\stackrel{4}{\circ}\rangle,
\end{align}
if we bring back to the standard order of 5-branes using the Hanany-Witten transitions and apply the relation $N_1+N_5=N_2+N_4$ in \eqref{constraint}.
Apparently, this brane configuration is not located in the original space \eqref{constraint} since $(N_1)+(N_5)\ne (N_2+2k)+(N_4+2k)$.
This is the reason we fix the labels for the two NS5-branes so that $\stackrel{\text{ii}}{\bullet}$ is on the left while $\stackrel{\text{i}}{\bullet}$ is on the right (corresponding to $\widehat{\cal P}_\text{ii}$ on the right and $\widehat{\cal P}_\text{i}$ on the left in terms of spectral operators).

Let us rewrite the action of the remaining generators $s_4$ and $s_3$ as brane transitions.
In terms of brane configurations, these generators act as
\begin{align}
&s_4:\langle N_1\stackrel{\text{ii}}{\bullet}N_2\stackrel{1}{\circ}N_3\stackrel{2}{\circ}
N_4\stackrel{\text{i}}{\bullet}N_5\stackrel{3}{\circ}N_6\stackrel{4}{\circ}\rangle\nonumber\\
&\mapsto\langle N_1\stackrel{\text{ii}}{\bullet}2N_1-N_2+2k\stackrel{1}{\circ}N_1-N_2+N_3+k\stackrel{2}{\circ}
N_4\stackrel{\text{i}}{\bullet}2N_4-N_5+2k\stackrel{3}{\circ}N_4-N_5+N_6+k\stackrel{4}{\circ}\rangle,
\nonumber\\
&s_3:\langle N_1\stackrel{\text{ii}}{\bullet}N_2\stackrel{1}{\circ}N_3\stackrel{2}{\circ}
N_4\stackrel{\text{i}}{\bullet}N_5\stackrel{3}{\circ}N_6\stackrel{4}{\circ}\rangle
\mapsto\langle N_1\stackrel{\text{ii}}{\bullet}N_3\stackrel{1}{\circ}N_2\stackrel{2}{\circ}
N_2-N_3+N_4\stackrel{\text{i}}{\bullet}N_5\stackrel{3}{\circ}N_6\stackrel{4}{\circ}\rangle.
\label{e7N}
\end{align}
Of course the expressions of these actions are ambiguous under the constraint $N_1+N_5=N_2+N_4$.
If we introduce auxiliary variables
\begin{align}
N'_2=N_1-N_2+N_3+k,\quad N'_5=N_4-N_5+N_6+k,\quad
N'_4=N_3-N_4+N_5+k,
\label{Nprime}
\end{align}
we can rewrite the transformations into the brane transitions
\begin{align}
s_4:\langle N_1\stackrel{1}{\circ}N'_2\stackrel{\text{ii}}{\bullet}N_3\stackrel{2}{\circ}
N_4\stackrel{3}{\circ}N'_5\stackrel{\text{i}}{\bullet}N_6\stackrel{4}{\circ}\rangle
\mapsto\langle N_1\stackrel{1}{\circ}N_3\stackrel{\text{ii}}{\bullet}N'_2\stackrel{2}{\circ}
N_4\stackrel{3}{\circ}N_6\stackrel{\text{i}}{\bullet}N'_5\stackrel{4}{\circ}\rangle,
\nonumber\\
s_3:\langle N_1\stackrel{\text{ii}}{\bullet}N_2\stackrel{1}{\circ}N_3\stackrel{\text{i}}{\bullet}
N'_4\stackrel{2}{\circ}N_5\stackrel{3}{\circ}N_6\stackrel{4}{\circ}\rangle
\mapsto\langle N_1\stackrel{\text{ii}}{\bullet}N_3\stackrel{1}{\circ}N_2\stackrel{\text{i}}{\bullet}
N'_4\stackrel{2}{\circ}N_5\stackrel{3}{\circ}N_6\stackrel{4}{\circ}\rangle.
\label{e7s4s3}
\end{align}
As previously, the $s_3$ transformation exchanges the numbers of D3-branes on two sides of the $(1,k)$5-brane $\stackrel{1}{\circ}$, while the $s_4$ transformation exchanges those on two sides of both the two NS5-branes $\stackrel{\text{ii}}{\bullet}$ and $\stackrel{\text{i}}{\bullet}$ simultaneously.

As reviewed in section \ref{d5}, in studying the $D_5$ quantum curve in \cite{KM}, we were able to extract the brane transitions \eqref{d5s3s4} unknown previously.
Correspondingly, we focus on the $E_7$ quantum curve in this section.
After separating $s_2s_5$, $s_1s_6$ and $s_0s_7$ as the Hanany-Witten transitions, we are able to extract $s_4$ and $s_3$ \eqref{e7N} as brane transitions unknown previously.
With the introduction of new variables in \eqref{Nprime}, these brane transitions are clearly expressed as \eqref{e7s4s3}.
We would like to present a cleaner interpretation for these brane transitions in the next section by comparing \eqref{d5s3s4} and \eqref{e7s4s3}.

\section{Implications}

Previously we have reviewed the brane transitions for the $D_5$ curve and proceeded further to the $E_7$ case by overcoming some difficulties originating from the degeneracies.
In this section, we would like to obtain a general implication from these two cases and provide some checks and discussions for it.

\subsection{Observations}

\begin{figure}[!t]
\centering\includegraphics[scale=0.5,angle=-90]{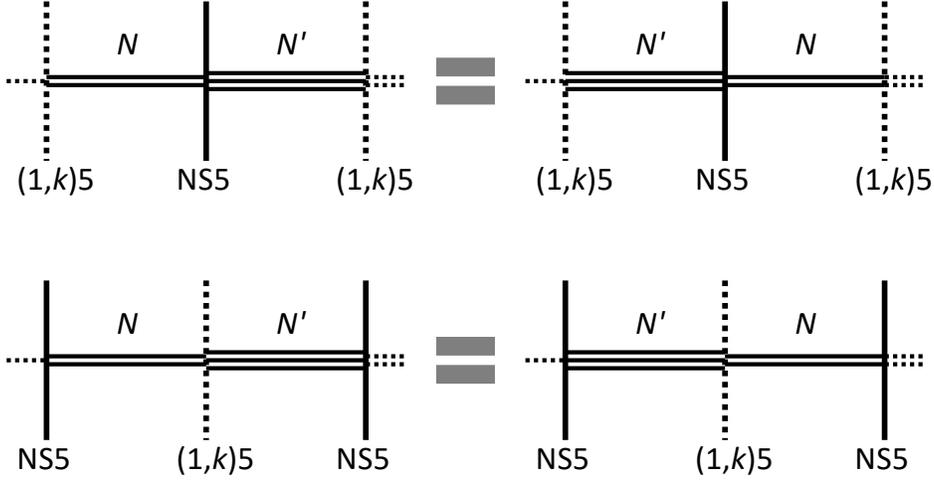}
\caption{Proposed local rule for the brane transitions.}
\label{localtransition}
\end{figure}

In the previous sections we have found that there are symmetries which cannot be understood from the established Hanany-Witten transitions.
Especially, we have identified them in terms of brane transitions in \eqref{d5N} and \eqref{e7N}.
After introducing new variables, we can simplify the transitions in \eqref{d5s3s4} and \eqref{e7s4s3}.
If there is a physical interpretation to understand these brane transitions, we may want to require the interpretation to be given ``locally''.
Here by locality we mean that the brane transitions should only relate to a small subset of branes instead of hold only after we consider the whole set of the brane configurations.
If we believe in this assumption, from \eqref{d5s3s4} and \eqref{e7s4s3}, we may want to propose the brane transitions
\begin{align}
\cdots\circ N\bullet N'\circ\cdots=\cdots\circ N'\bullet N\circ\cdots,\quad
\cdots\bullet N\circ N'\bullet\cdots=\cdots\bullet N'\circ N\bullet\cdots,
\label{local}
\end{align}
where $\cdots$ can be any array of D3-branes connected by any type of 5-branes as long as they are unchanged between both sides of the equations.
See figure \ref{localtransition} for a brane picture of the transitions.

Physically, it is surprising for us to observe these brane transitions.
Since the ordering of D3-branes may change the brane dynamics drastically, we believe that the brane transitions should hold only restrictively.
It is of course an interesting future direction to specify the situations when the brane transitions \eqref{local} hold.
Especially, it is crucial to understand whether the transitions are accidental for curves of genus one since their origins are the exceptional Weyl groups characteristic for these curves.
In the following subsections, we shall provide some checks and discussions for the proposal \eqref{local}.

\subsection{$D_5$ curve}

If the proposed transitions \eqref{local} hold for the $D_5$ case, we definitely also want to require symmetries such as those transforming $\langle N_1\stackrel{2}{\bullet}N_2\stackrel{3}{\circ}N'_3\stackrel{1}{\bullet}N_4\stackrel{4}{\circ}\rangle$ (with $N'_3=N_2-N_3+N_4+k$) into $\langle N_4\stackrel{2}{\bullet}N_2\stackrel{3}{\circ}N'_3\stackrel{1}{\bullet}N_1\stackrel{4}{\circ}\rangle$ or $\langle N_2\stackrel{2}{\bullet}N_1\stackrel{3}{\circ}N'_3\stackrel{1}{\bullet}N_4\stackrel{4}{\circ}\rangle$ by applying the transitions \eqref{local} to different parts.
Since these transformations do not preserve the reference rank $N_1$, we need to bring the reference rank back to $N_1$ by rotating the brane configurations cyclically as in
\begin{align}
&\langle N_4\stackrel{2}{\bullet}N_2\stackrel{3}{\circ}N'_3\stackrel{1}{\bullet}N_1\stackrel{4}{\circ}\rangle\to
\langle N_1\stackrel{4}{\circ}N_4\stackrel{2}{\bullet}N_2\stackrel{3}{\circ}N'_3\stackrel{1}{\bullet}\rangle,
\nonumber\\
&\langle N_2\stackrel{2}{\bullet}N_1\stackrel{3}{\circ}N'_3\stackrel{1}{\bullet}N_4\stackrel{4}{\circ}\rangle\to
\langle N_1\stackrel{3}{\circ}N'_3\stackrel{1}{\bullet}N_4\stackrel{4}{\circ}N_2\stackrel{2}{\bullet}\rangle.
\end{align}
If we bring the 5-branes into the original standard order using the Hanany-Witten transitions \eqref{HW}, the transformations map $\langle N_1\stackrel{2}{\bullet}N_2\stackrel{1}{\bullet}N_3\stackrel{3}{\circ}N_4\stackrel{4}{\circ}\rangle$ into
\begin{align}
\langle N_1\stackrel{2}{\bullet}N_1+N_2-N_4+k\stackrel{1}{\bullet}2N_1+N_3-2N_4+2k\stackrel{3}{\circ}2N_1-N_4+2k\stackrel{4}{\circ}\rangle,
\nonumber\\
\langle N_1\stackrel{2}{\bullet}2N_1-N_2+2k\stackrel{1}{\bullet}2N_1-2N_2+N_3+2k\stackrel{3}{\circ}N_1-N_2+N_4+k\stackrel{4}{\circ}\rangle.
\end{align}
We can rewrite the resulting transformations in terms of the rank differences $(M_1,M_2,M_3)$ introduced in \eqref{M1M2M3} as
\begin{align}
&(M_1,M_2,M_3)\mapsto(M_2+M_3+k,(M_1+M_2-M_3-k)/2,(M_1-M_2+M_3-k)/2),\nonumber\\
&(M_1,M_2,M_3)\mapsto(M_2-M_3+k,(M_1+M_2+M_3-k)/2,(-M_1+M_2+M_3+k)/2).
\label{m123transf}
\end{align}
Furthermore, using \eqref{hef}, we can express the transformations in terms of $(m_1,m_2,m_3)$.

From the viewpoint of quantum curves we already know that the symmetry for the three-dimensional space of brane configurations ${\cal C}_\text{B}^{2,2}$ is given by $W(B_3)$.
For the consistency of the proposed brane transitions \eqref{local}, the transformations \eqref{m123transf} have to be elements of $W(B_3)$.
Here we find that in fact these transformations are generated in $W(B_3)$ as
\begin{align}
(s_1s_2)s_3(s_1s_2)&:(m_1,m_2,m_3)\mapsto(m_2m_3,(m_1m_2m_3^{-1})^{\frac{1}{2}},(m_1m_2^{-1}m_3)^{\frac{1}{2}}),\nonumber\\
(s_5s_0)s_4(s_5s_0)&:(m_1,m_2,m_3)\mapsto(m_2m_3^{-1},(m_1m_2m_3)^{\frac{1}{2}},(m_1^{-1}m_2m_3)^{\frac{1}{2}}).
\end{align}

Interestingly, there is a simple interpretation for the above elements of the Weyl group $W(B_3)$.
Originally in \eqref{d5s3s4}, the $s_3$ transformation exchanges the numbers of D3-branes on two sides of the $(1,k)$5-brane $\stackrel{3}{\circ}$, while the $s_4$ transformation exchanges those on two sides of the NS5-brane $\stackrel{1}{\bullet}$.
Here to apply the transitions \eqref{local} to the $(1,k)$5-brane $\stackrel{4}{\circ}$ and the NS5-brane $\stackrel{2}{\bullet}$, we need to swap $\stackrel{3}{\circ}$ for $\stackrel{4}{\circ}$ and $\stackrel{1}{\bullet}$ for $\stackrel{2}{\bullet}$ respectively.
These are generated respectively by conjugations of $s_3$ and $s_4$ by the Hanany-Witten transitions $s_1s_2$ and $s_5s_0$.

In the previous subsection, we have proposed the local brane transitions \eqref{local} from the results in \eqref{d5s3s4} and \eqref{e7s4s3}.
To confirm the proposal we have applied the transitions to different parts in the brane configurations.
We have found that in fact these brane transitions are elements of $W(B_3)$ and identified them as $(s_1s_2)s_3(s_1s_2)$ and $(s_5s_0)s_4(s_5s_0)$ respectively.
Hence, the brane configurations obtained from \eqref{local} are indeed symmetries, which we regard as an evidence for our proposal.
We have to confess that this consistency check may not be fully non-trivial, since they are simply generated from the brane transitions of $s_3$ or $s_4$ along with some trivial conjugations.
The best way to provide a more non-trivial evidence is to repeat in a different setup.
In the next subsection, we shall move to the brane system with two NS5-branes and four $(1,k)$5-branes studied in section \ref{e7}.

\subsection{$E_7$ curve}

In the previous subsections, we have guessed the brane transitions \eqref{local} and checked the consistency for the $D_5$ curve.
For the $E_7$ case the resulting brane transitions \eqref{e7s4s3} seem to support this guess in the subspace subject to the constraint \eqref{constraint}.
We explain this interpretation and continue to provide some checks for \eqref{local} in this subsection.

If the brane transitions \eqref{local} hold, we should be able to exchange the numbers of D3-branes on two sides of the NS5-brane when they are surrounded by two $(1,k)$5-branes.
This implies that we could in principle transform $\langle N_1\stackrel{1}{\circ}N'_2\stackrel{\text{ii}}{\bullet}N_3\stackrel{2}{\circ}N_4\stackrel{3}{\circ}N'_5\stackrel{\text{i}}{\bullet}N_6\stackrel{4}{\circ}\rangle$ into $\langle N_1\stackrel{1}{\circ}N_3\stackrel{\text{ii}}{\bullet}N'_2\stackrel{2}{\circ}N_4\stackrel{3}{\circ}N'_5\stackrel{\text{i}}{\bullet}N_6\stackrel{4}{\circ}\rangle$ or $\langle N_1\stackrel{1}{\circ}N'_2\stackrel{\text{ii}}{\bullet}N_3\stackrel{2}{\circ}N_4\stackrel{3}{\circ}N_6\stackrel{\text{i}}{\bullet}N'_5\stackrel{4}{\circ}\rangle$.
However this would break the constraint \eqref{constraint}.
Due to this reason, to preserve the constraint we need to apply the transitions \eqref{local} to two sides of both the NS5-branes $\stackrel{\text{ii}}{\bullet}$ and $\stackrel{\text{i}}{\bullet}$ simultaneously in the $s_4$ transformation \eqref{e7s4s3}.

Similarly for $s_3$ in \eqref{e7s4s3}, when we bring the NS5-brane $\stackrel{\text{i}}{\bullet}$ forward by applying the Hanany-Witten transition \eqref{HW}, so that the NS5-brane $\stackrel{\text{i}}{\bullet}$ is closer to $\stackrel{\text{ii}}{\bullet}$ and the $(1,k)$5-brane $\stackrel{1}{\circ}$ is surrounded by two NS5-branes in $\langle N_1\stackrel{\text{ii}}{\bullet}N_2\stackrel{1}{\circ}N_3\stackrel{\text{i}}{\bullet}N'_4\stackrel{2}{\circ}N_5\stackrel{3}{\circ}N_6\stackrel{4}{\circ}\rangle$, we are able to exchange the numbers of D3-branes on two sides of the $(1,k)$5-brane $\stackrel{1}{\circ}$ without breaking the constraint \eqref{different1k5}.

Instead of bringing the NS5-brane $\stackrel{\text{i}}{\bullet}$ forward in the $s_3$ transformation, we can alternatively send it backward so that the brane configuration is $\langle N_1\stackrel{\text{ii}}{\bullet}N_2\stackrel{1}{\circ}N_3\stackrel{2}{\circ}N_4\stackrel{3}{\circ}N'_5\stackrel{\text{i}}{\bullet}N_6\stackrel{4}{\circ}\rangle$ with $N'_5=N_4-N_5+N_6+k$.
Then we can exchange the numbers of D3-branes on two sides of the $(1,k)$5-brane $\stackrel{4}{\circ}$ across the reference rank $N_1$, $\langle N_6\stackrel{\text{ii}}{\bullet}N_2\stackrel{1}{\circ}N_3\stackrel{2}{\circ}N_4\stackrel{3}{\circ}N'_5\stackrel{\text{i}}{\bullet}N_1\stackrel{4}{\circ}\rangle$.
To keep the reference rank $N_1$, we cyclically rotate the brane configuration and obtain $\langle N_1\stackrel{4}{\circ}N_6\stackrel{\text{ii}}{\bullet}N_2\stackrel{1}{\circ}N_3\stackrel{2}{\circ}N_4\stackrel{3}{\circ}N'_5\stackrel{\text{i}}{\bullet}\rangle$.
After applying the Hanany-Witten transitions to bring the 5-branes back to the standard order, we find
\begin{align}
&\langle N_1\stackrel{4}{\circ}N_6\stackrel{\text{ii}}{\bullet}N_2\stackrel{1}{\circ}
N_3\stackrel{2}{\circ}N_4\stackrel{3}{\circ}N'_5\stackrel{\text{i}}{\bullet}\rangle
\nonumber\\
&=\langle N_1\stackrel{\text{ii}}{\bullet}N_2+N'\stackrel{1}{\circ}N_3+N'\stackrel{2}{\circ}
N_4+N'\stackrel{\text{i}}{\bullet}N_5+2N'\stackrel{3}{\circ}N_6+2N'\stackrel{4}{\circ}\rangle,
\label{4circ}
\end{align}
with $N'=N_1-N_6+k$, which respects the constraint \eqref{constraint}.
After rewriting the transformation in terms of parameters $(f_1,f_2,f_3,g_1)$ of ${\cal C}_\text{B}^{2,4}\cap{\cal C}_\text{P}^{E_7}$ \eqref{e7sub} using \eqref{FGN}, we find the transformation is given by
\begin{align}
(s_0s_7)(s_1s_6)(s_2s_5)s_3(s_2s_5)(s_1s_6)(s_0s_7)&:(f_1,f_2,f_3,g_1)\nonumber\\
&\mapsto\biggl(\sqrt{\frac{f_1g_1}{f_2f_3}},\sqrt{\frac{f_2g_1}{f_1f_3}},\sqrt{\frac{f_3g_1}{f_1f_2}},\sqrt{f_1f_2f_3g_1}\biggr),
\label{0716253}
\end{align}
which exists in the Weyl group $W(F_4)$.

Similarly, we can send the NS5-brane $\stackrel{\text{ii}}{\bullet}$ backward so that the brane configuration is $\langle N_1\stackrel{1}{\circ}N_2'\stackrel{\text{ii}}{\bullet}N_3\stackrel{2}{\circ}N_4\stackrel{\text{i}}{\bullet}N_5\stackrel{3}{\circ}N_6\stackrel{4}{\circ}\rangle$ with $N'_2=N_1-N_2+N_3+k$.
Then after exchanging the numbers of D3-branes on two sides of the $(1,k)$5-brane $\stackrel{2}{\circ}$ and bringing back to the standard order, we find the brane configuration
\begin{align}
\langle N_1\stackrel{1}{\circ}N_2'\stackrel{\text{ii}}{\bullet}N_4\stackrel{2}{\circ}N_3\stackrel{\text{i}}{\bullet}N_5\stackrel{3}{\circ}N_6\stackrel{4}{\circ}\rangle
=\langle N_1\stackrel{\text{ii}}{\bullet}N_2-N_3+N_4\stackrel{1}{\circ}N_4\stackrel{2}{\circ}
N_3\stackrel{\text{i}}{\bullet}N_5\stackrel{3}{\circ}N_6\stackrel{4}{\circ}\rangle.
\label{2circ}
\end{align}
Again this transformation exists in the Weyl group $W(F_4)$,
\begin{align}
(s_2s_5)s_3(s_2s_5):(f_1,f_2,f_3,g_1)\mapsto\biggl(f_1,\sqrt{\frac{f_1f_3}{f_2g_1}},f_3,\sqrt{\frac{f_1f_3g_1}{f_2^3}}\biggr).
\label{253}
\end{align}

As previously, the elements of $W(F_4)$ have the following simple interpretation.
Since the original $s_3$ transformation in \eqref{e7s4s3} exchanges the numbers of D3-branes on two sides of the $(1,k)$5-brane $\stackrel{1}{\circ}$, if we apply the transition \eqref{local} to the $(1,k)$5-brane $\stackrel{4}{\circ}$ in \eqref{4circ}, we need to conjugate $s_3$ by a series of the Hanany-Witten transitions $s_2s_5$, $s_1s_6$, $s_0s_7$ all the way to exchange $\stackrel{1}{\circ}$/$\stackrel{2}{\circ}$, $\stackrel{2}{\circ}$/$\stackrel{3}{\circ}$ and finally $\stackrel{3}{\circ}$/$\stackrel{4}{\circ}$ in \eqref{0716253}.
Also, if we apply the transition \eqref{local} to the $(1,k)$5-brane $\stackrel{2}{\circ}$ in \eqref{2circ}, we can conjugate $s_3$ simply by $s_2s_5$ in \eqref{253}.

As a slightly different check, we can perform the following computation.
We can apply the transitions \eqref{local} to both the NS5-branes $\bullet$ in the original brane configuration $\langle N_1\stackrel{\text{ii}}{\bullet}N_2\stackrel{1}{\circ}N_3\stackrel{2}{\circ}N_4\stackrel{\text{i}}{\bullet}N_5\stackrel{3}{\circ}N_6\stackrel{4}{\circ}\rangle$ without breaking the constraint \eqref{constraint}.
As a result, we find $\langle N_2\stackrel{\text{ii}}{\bullet}N_1\stackrel{1}{\circ}N_3\stackrel{2}{\circ}N_5\stackrel{\text{i}}{\bullet}N_4\stackrel{3}{\circ}N_6\stackrel{4}{\circ}\rangle$, which reduces to $\langle N_1\stackrel{1}{\circ}N_3\stackrel{2}{\circ}N_5\stackrel{\text{i}}{\bullet}N_4\stackrel{3}{\circ}N_6\stackrel{4}{\circ}N_2\stackrel{\text{ii}}{\bullet}\rangle$ after rotating cyclically.
The rotation, however, changes the order of the two NS5-branes, and breaks the constraint \eqref{constraint}.
To return to the original subspace ${\cal C}_\text{B}^{2,4}\cap{\cal C}_\text{P}^{E_7}$, we need to exchange the two NS5-branes,
\begin{align}
&\langle N_1\stackrel{1}{\circ}N_3\stackrel{2}{\circ}N_5\stackrel{\text{ii}}{\bullet}N_4\stackrel{3}{\circ}N_6\stackrel{4}{\circ}N_2\stackrel{\text{i}}{\bullet}\rangle
\nonumber\\
&=\langle N_1\stackrel{\text{ii}}{\bullet}N_2+2N'\stackrel{1}{\circ}N_3+N'\stackrel{2}{\circ}N_4\stackrel{\text{i}}{\bullet}N_5+2N'\stackrel{3}{\circ}N_6+N'\stackrel{4}{\circ}\rangle,
\end{align}
with
\begin{align}
N'=N_1-N_2+k=N_4-N_5+k.
\end{align}
It is interesting to see whether this transformation still remains in $W(F_4)$.
Surprisingly, it turns out that the transformation is nothing but the original transformation $s_4$ \eqref{e7s4s3}.
This consistency check is different from previous ones in the sense that the brane configurations do not always reside in ${\cal C}_\text{P}^{E_7}$.
Even though we have evaded the original subspace ${\cal C}_\text{B}^{2,4}\cap{\cal C}_\text{P}^{E_7}$ once, the transformation still exists in $W(F_4)$.

Here we have regarded the fact that the Weyl group $W(F_4)$ contains the symmetries expected from the proposed brane transitions \eqref{local} as an evidence for the proposal.
As we stress previously, this may still not be fully non-trivial.
We wish to have more non-trivial checks for our proposal.
In the next subsection, we stop searching for evidences for the proposal \eqref{local} and turn to discuss an implication of it.

\subsection{Beyond $E_7$ curve}\label{infinite}

So far we have presented several checks for the brane transitions \eqref{local}.
Let us turn to discuss the implications.
Concretely, let us assume that the brane transitions \eqref{local} are applicable to the whole space of brane configurations ${\cal C}_\text{B}^{2,4}$ instead of its subspace ${\cal C}_\text{B}^{2,4}\cap{\cal C}_\text{P}^{E_7}$ to find out a consequence of the brane transitions \eqref{local}.
Note of course that, since we are applying the transitions \eqref{local} obtained within ${\cal C}_\text{P}^{E_7}$ to those outside ${\cal C}_\text{P}^{E_7}$, the computation should be interpreted with caution.

In the discussions of the $s_3$ transformation for the $E_7$ curve, we have brought the NS5-brane $\stackrel{\text{i}}{\bullet}$ forward in $\langle N_1\stackrel{\text{ii}}{\bullet}N_2\stackrel{1}{\circ}N_3\stackrel{\text{i}}{\bullet}N'_4\stackrel{2}{\circ}N_5\stackrel{3}{\circ}N_6\stackrel{4}{\circ}\rangle$ (with $N'_4=N_3-N_4+N_5+k$) satisfying the constraint $N_1+N'_4=N_2+N_3+k$ \eqref{different1k5}.
Instead of applying the brane transition \eqref{local} by exchanging the numbers of D3-branes on two sides of the $(1,k)$5-brane $\stackrel{1}{\circ}$, let us consider the application of the brane transition \eqref{local} by exchanging the numbers on two sides of both the NS5-branes $\stackrel{\text{i}}{\bullet}$ and $\stackrel{\text{ii}}{\bullet}$ simultaneously and obtain $\langle N_2\stackrel{\text{ii}}{\bullet}N_1\stackrel{1}{\circ}N'_4\stackrel{\text{i}}{\bullet}N_3\stackrel{2}{\circ}N_5\stackrel{3}{\circ}N_6\stackrel{4}{\circ}\rangle$.
If we return to the standard order of 5-branes with the Hanany-Witten transition, what we obtain is $\langle N_2\stackrel{\text{ii}}{\bullet}N_1\stackrel{1}{\circ}N'_4\stackrel{2}{\circ}N''_3\stackrel{\text{i}}{\bullet}N_5\stackrel{3}{\circ}N_6\stackrel{4}{\circ}\rangle$ with $N ''_3=N'_4-N_3+N_5+k$.
Let us apply the transition \eqref{local} again to exchange the numbers on two sides of both the NS5-branes $\stackrel{\text{i}}{\bullet}$ and $\stackrel{\text{ii}}{\bullet}$ simultaneously and obtain $\langle N_1\stackrel{\text{ii}}{\bullet}N_2\stackrel{1}{\circ}N'_4\stackrel{2}{\circ}N_5\stackrel{\text{i}}{\bullet}N''_3\stackrel{3}{\circ}N_6\stackrel{4}{\circ}\rangle$.
This brane configuration does not reside in the original subspace ${\cal C}_\text{B}^{2,4}\cap{\cal C}_\text{P}^{E_7}$ any more and the constraint turns out to be
\begin{align}
\{\langle N_1\stackrel{\text{ii}}{\bullet}N_2\stackrel{1}{\circ}N_3\stackrel{2}{\circ}
N_4\stackrel{\text{i}}{\bullet}N_5\stackrel{3}{\circ}N_6\stackrel{4}{\circ}\rangle|N_1+N_5=N_2+N_4+2k\}.
\end{align}
Namely after these brane transitions the original subspace ${\cal C}_\text{B}^{2,4}\cap{\cal C}_\text{P}^{E_7}$ is shifted by $2k$ in parallel.
If we repeat these brane transitions for multiple times, the subspace will continue to be shifted.
This implies that the brane transitions \eqref{local} generate symmetries of infinite dimensions in general.

From the viewpoint of the quantum curves, this may not be so surprising.
As stressed in \eqref{constraint}, the space of brane configurations ${\cal C}_\text{B}^{2,4}$ is not located in ${\cal C}_\text{P}^{E_7}$ and the corresponding curve is not of genus one but of genus three in general as can be read off from the toric diagram.
Considering that only curves of genus one enjoy the symmetries given by finite-dimensional Weyl groups of exceptional groups, it is natural to encounter the infinite-dimensional symmetries for the space of brane configurations ${\cal C}_\text{B}^{2,4}$.
Hence, we believe that the consequence of the infinite dimensions holds even though the proposed transitions can be modified finally.

\section{Conclusion}

Previously in \cite{KM} by embedding the space of brane configurations ${\cal C}_\text{B}^{2,2}$ into the space of point configurations ${\cal C}_\text{P}^{D_5}$, new brane transitions were found from the $D_5$ Weyl group.
Aiming at understanding the brane transitions better, in this paper we generalize the previous studies to the $E_7$ quantum curve.
Although the situation is quite different due to degeneracies of the curve, we can still embed the space of brane configurations ${\cal C}_\text{B}^{2,4}$ partially into the space of point configurations ${\cal C}_\text{P}^{E_7}$ and find similar new brane transitions from the $E_7$ Weyl group as well.
Since both the new brane transitions found for the $D_5$ curve and the $E_7$ curve resemble each other, we try to propose a ``local'' rule, which does not refer to branes at a distance.
We present some checks for the proposed transitions, though this may not be fully non-trivial.
We list several future directions in the following we would like to pursue.

Firstly, it is of course an interesting future direction to collect more non-trivial evidences for the proposal \eqref{local} or even verify it.
Here we have checked that the proposal is consistent with the Weyl group, though we believe that this is not enough.
Recent studies of the $\text{SL}(2,{\mathbb Z})$ transformation \cite{A} or the spectral operators with rank deformations \cite{K} generalized from \cite{KMZ} may be helpful in non-trivial checks of our proposal.

Secondly, although we have only focused on the studies of partition functions, the Hanany-Witten transitions are known to be applicable to other excitations in the field theories.
It is natural to ask whether the brane transitions proposed in this paper work for those excitations.
It is especially interesting to start with the half-BPS Wilson loop operators \cite{HHMO,2PT,closed}.

Thirdly, to utilize Weyl groups of exceptional groups, we have focused mainly on the del Pezzo geometries of genus one.
In section \ref{infinite}, we have considered more general setups to discuss the implications of the proposed brane transitions and find symmetries of infinite dimensions.
To discuss the infinite-dimensional symmetries for these setups, it is important to extend the studies of Fredholm determinants to higher genus in \cite{HM,MN3,HHO} or \cite{CGM,CGuM,HSW}.

Fourthly, it is known that Weyl groups of exceptional groups studied in this paper play an important role in the Painlev\'e equations.
It is interesting to investigate a direct relation between the brane configurations and the Painlev\'e equations.
This may lead us to understand the relations between the ABJM matrix model and the Painlev\'e equation \cite{GHM2,BGT}, the integrable hierarchy \cite{MM,G,JT,2DTL}, the chiral projection \cite{H,MS2,MN5} or the TBA system \cite{HMMO,OZ,DH} more systematically.

\section*{Acknowledgements}
We are grateful to Naotaka Kubo, Tomoki Nosaka, Yutaka Ookouchi, Yuji Sugimoto, Yasuhiko Yamada and Katsuya Yano for valuable discussions.
The work of T.F.\ and S.M.\ is supported respectively by Grant-in-Aid for JSPS Fellows \#20J15045 and Grant-in-Aid for Scientific Research (C) \#19K03829.
S.M.\ would like to thank Yukawa Institute for Theoretical Physics at Kyoto University for warm hospitality.

\end{document}